\documentclass[conference]{IEEEtran}
\IEEEoverridecommandlockouts
\usepackage{amsmath,amssymb,amsfonts}
\usepackage{braket}
\usepackage{acro}
\usepackage{graphicx}
\usepackage{subcaption}
\usepackage{color, soul}

\def\BibTeX{{\rm B\kern-.05em{\sc i\kern-.025em b}\kern-.08em
    T\kern-.1667em\lower.7ex\hbox{E}\kern-.125emX}}

\DeclareAcronym{crb}{short=CRB, long=Clifford randomized benchmarking}
\DeclareAcronym{epcl}{short=EPCL, long=Error Per Circuit Layer}
\DeclareAcronym{drb}{short=DRB, long=direct randomized benchmarking}
\DeclareAcronym{rb}{short=RB, long=randomized benchmarking}
\DeclareAcronym{mrb}{short=MRB, long=mirror randomized benchmarking}
\DeclareAcronym{birb}{short=BiRB, long=binary randomized benchmarking}
\DeclareAcronym{irb}{short=IRB, long=interleaved randomized benchmarking}
\DeclareAcronym{eplg}{short=EPLG, long=error per layered gate}
\DeclareAcronym{spam}{short=SPAM, long=state preparation and measurement}
\DeclareAcronym{xeb}{short=XEB, long=cross-entropy benchmarking}
\DeclareAcronym{qpu}{short=QPU, long=quantum processing unit}
\DeclareAcronym{qec}{short=QEC, long=quantum error correction}
\DeclareAcronym{rmse}{short=RMSE, long=root mean square error}

\AtBeginDocument{%
  \providecommand\BibTeX{{%
    Bib\TeX}}}

\begin{document}

\makeatletter
\newcommand{\linebreakand}{%
  \end{@IEEEauthorhalign}
  \hfill\mbox{}\par
  \mbox{}\hfill\begin{@IEEEauthorhalign}
}
\makeatother

\title{Characterizing Large Scale Quantum Systems with Error Per Circuit Layer}

\author{
    \IEEEauthorblockN{
        Travis Hurant\IEEEauthorrefmark{1}\IEEEauthorrefmark{2}, 
        Arian Vezvaee\IEEEauthorrefmark{3}\IEEEauthorrefmark{4}, 
        Swarnadeep Majumder\IEEEauthorrefmark{1}\IEEEauthorrefmark{2}$^1$, 
        Aniket Dalvi\IEEEauthorrefmark{1}\IEEEauthorrefmark{2}$^2$, 
        Jude Alnas\IEEEauthorrefmark{1}\IEEEauthorrefmark{2},  
        \linebreakand
        Kenneth R. Brown\IEEEauthorrefmark{1}\IEEEauthorrefmark{2}
        \IEEEauthorrefmark{8}\IEEEauthorrefmark{9}
    }
    \IEEEauthorblockA{\IEEEauthorrefmark{1} Duke Quantum Center, Duke University, Durham, NC USA}
    \IEEEauthorblockA{\IEEEauthorrefmark{2} Department of Electrical and Computer Engineering, Duke University, Durham, NC USA}
    \IEEEauthorblockA{\IEEEauthorrefmark{3} Department of Electrical and Computer Engineering, University of Southern California, Los Angeles, CA USA}
    \IEEEauthorblockA{\IEEEauthorrefmark{4} Center for Quantum Information Science and Technology, University of Southern California, Los Angeles, CA USA}
    \IEEEauthorblockA{\IEEEauthorrefmark{8} Department of Physics, Duke University, Durham, NC USA}
    \IEEEauthorblockA{\IEEEauthorrefmark{9} Department of Chemistry, Duke University, Durham, NC USA}
    \thanks{$^1$Swarnadeep Majumder is currently affiliated with IBM. 
    
    $^2$Aniket Dalvi is currently affiliated with Amazon Web Services (AWS).}
}

\maketitle

\begin{abstract}
Quantum benchmarks provide compact measures of performance that are important for evaluating and comparing quantum systems. Circuit-level benchmarks are particularly valuable because they capture the accumulated effects of noise across interacting operations, but existing approaches may require structured gate sets and costly compilation, classical simulation of reference outputs, or subsystem decompositions that do not capture full-register behavior. We introduce \ac{epcl}, an overlap-based circuit-level benchmark that estimates an effective layer polarization by applying identical random circuits to two disjoint quantum registers and measuring the overlap between their output states as a function of circuit depth. \ac{epcl} avoids classical simulation of ideal output distributions and recovery to a known reference state, and is compatible with arbitrary gate sets, including non-Clifford gates. 

We derive the expected overlap decay under an ensemble-averaged depolarizing model and identify the assumptions under which the fitted decay parameter represents an effective layer polarization. Numerical simulations show that \ac{epcl} recovers the predicted polarization under weak local stochastic noise and remains well described by a single-exponential decay at stronger stochastic noise levels. The simulations further show that coherent errors associated with fixed entangling layers may require Pauli twirling or randomized compiling to produce the expected decay, while inter-register correlations contribute an additional covariance term to the measured overlap. Finally, experiments on IBM quantum hardware demonstrate clear \ac{epcl} decay in 8- and 16-qubit implementations. These results support \ac{epcl} as a method for measuring aggregate register performance without requiring classical simulation of ideal circuit outputs or restriction to structured gate sets.

\end{abstract}

\section*{Introduction}

Quantum benchmarks are a critical component of the quantum computing workflow. The diagnostic information they provide informs calibration schedules, execution strategies, and whether a device is operating within performance regimes required for fault-tolerant quantum computation. Benchmarking protocols are often categorized as gate-level, circuit-level, or processor-level according to the scale at which they probe system performance \cite{proctor2025benchmarking}. While gate-level techniques characterize individual operations and processor-level benchmarks assess application performance, circuit-level benchmarks occupy an important middle ground by capturing the cumulative effects of noise across many interacting operations, including error mechanisms that may be difficult to infer from gate-level characterization alone.

Existing circuit-level benchmarks include \ac{mrb} \cite{proctor2022scalable}, \ac{xeb} \cite{Boixo2018}, and \ac{eplg} \cite{mckay2023benchmarking}. Each has demonstrated utility in this intermediate regime, but each also retains important limitations. For example, mirror randomized benchmarking relies on structured gate sets, cross-entropy benchmarking requires access to classically computed reference quantities, and error per-layered gate partitions the system into subsystems, which may reduce sensitivity to error processes that span those partitions. These limitations underscore the difficulty of designing circuit-level benchmarks that support universal gate sets, scale to increasingly large quantum systems, and remain sensitive to correlated error processes that extend beyond local subsystems. This need is particularly relevant in modular and error-corrected architectures, where complete registers or code blocks form natural computational units and their aggregate performance and mutual isolation become important system-level properties \cite{Bluvstein_2026,Monroe_2014}.

In this work, we introduce Error Per Circuit Layer (EPCL), a circuit-level benchmarking protocol designed to balance these competing considerations. \Ac{epcl} estimates an effective layer polarization from overlap measurements between two identically evolved quantum registers, avoiding the need for classical simulation of ideal output distributions. The protocol is compatible with arbitrary, including non-Clifford, gate sets and does not require inversion procedures or specialized state preparation. By applying randomized circuit layers to circuit-level subsystems, \ac{epcl} captures the accumulated effect of noise across many interacting operations and enables the investigation of correlated error processes that span multiple qubits or subsystems.

To establish the behavior and applicability of the protocol, we derive the expected \ac{epcl} overlap decay under an ensemble-averaged depolarizing model and show that the resulting observable estimates an effective layer polarization. Numerical simulations recover the expected polarization under weak local stochastic noise and show that the characteristic exponential decay remains well behaved beyond the regime in which a perturbative gate-level description is accurate. The simulations further identify important interpretive boundaries, including the need to randomize fixed coherent errors and the contribution of inter-register correlations to the measured overlap. Finally, experiments on IBM quantum hardware demonstrate clear \ac{epcl} decay using both two $4$-qubit and two $8$-qubit registers, including a $16$-qubit implementation under realistic device noise.

\section{Background}

\Ac{epcl} incorporates aspects of \ac{rb} and \ac{xeb}, but utilizes an overlap-based observable to estimate circuit layer fidelity. In this section we briefly describe these foundational components.

\subsection{Randomized Benchmarking}

\Ac{rb} comprises a family of protocols that estimate average gate performance using sequences of random circuits of varying depth \cite{Emerson_2005, Knill_2008}. In general, \ac{rb} protocols are robust to \ac{spam} errors and scale favorably compared to tomographic techniques \cite{magesan_2012}. We briefly describe \ac{crb} \cite{Magesan_2011} as it is commonly regarded as the canonical form of randomized benchmarking and serves as the foundation for many later variants. 

In \ac{crb}, sequences of randomly sampled Clifford gates are applied to an $n$-qubit system, followed by a final inversion gate chosen such that, in the absence of noise, the system is returned to its initial state. In the presence of noise, the probability of recovering the initial state decays as a function of circuit depth. Averaging over random circuit instances yields a decay curve that is well-approximated by an exponential of the form \cite{magesan_2012}
\begin{align}
    \bar{P}(m) = Af^m + B,
\end{align}
where $A$ and $B$ capture SPAM contributions and $f$ is interpreted as an estimate of the average process polarization. The corresponding average Clifford gate infidelity is given by
\begin{align}
    r = \frac{2^n - 1}{2^n}(1-f).
\end{align}

Although \ac{rb} has become a standard tool in quantum benchmarking, its reliance on specific algebraic structures, most commonly the Clifford group, can limit scalability and applicability \cite{Cross_2016}. In multi-qubit settings, variants such as \ac{crb} become increasingly difficult to apply at larger system sizes because random $n$-qubit Clifford operations require substantial classical compilation overhead and deep native-gate decompositions \cite{Proctor_2019}. As a result, practical implementations are typically limited to relatively small subsystems. Moreover, the theoretical guarantees underlying many of these protocols depend on the availability of suitable group structures and are often tied to metrics based on recovery of a known reference state. These limitations motivate the exploration of complementary approaches that probe more general circuit behavior on larger subsystems.

\subsection{Cross Entropy Benchmarking}

\Ac{xeb} evaluates quantum system performance by comparing experimentally observed output distributions to those of ideal random circuits. In contrast to RB-style protocols, \ac{xeb} operates on ensembles of sufficiently random, non-Clifford circuits, enabling the characterization of system behavior beyond structured gate sets \cite{Arute2019, Boixo2018}.

In \ac{xeb}, the noisy output of a random circuit $U$ is modeled as a mixture of the ideal output state and an error component,
\begin{align}
    \rho_U = F \ket{\psi_U}\!\bra{\psi_U} + (1-F) \chi_U ,
\end{align}
where $F$ is the circuit fidelity and $\chi_U$ captures the contribution from erroneous circuit realizations. For sufficiently random circuits, the error component is assumed to be largely uncorrelated with the ideal output-probability “speckle” pattern \cite{Arute2019}. Under Porter-Thomas statistics, this leads to the linear-XEB estimator \cite{Arute2019, Boixo2018}
\begin{align}
    F_{\mathrm{XEB}} = \langle D p_s(q) - 1 \rangle ,
\end{align}
where $D=2^n$ and $p_s(q)$ is the ideal simulated probability of the measured bitstring $q$. Averaged over an ensemble of random circuits, the resulting \ac{xeb} fidelity exhibits an exponential decay with circuit depth. This decay is commonly interpreted in terms of an effective process polarization, analogous to randomized benchmarking. In this work, we adopt this model as the basis for our theoretical analysis.

However, \ac{xeb} requires knowledge of the ideal output probabilities for each circuit instance, which must be obtained through classical simulation. As system size increases, this requirement becomes intractable, limiting the applicability of \ac{xeb} to regimes where ideal output distributions remain classically accessible \cite{Boixo2018, Hashim_2021}. This motivates the search for observables that retain the favorable statistical properties of random-circuit benchmarking while avoiding the need for classical simulation.

\subsection{Overlap-Based Observables}

In contrast to benchmarking protocols based on output probabilities or state recovery, EPCL considers the overlap between two quantum states, defined as
\begin{align}
    \mathrm{Tr}(\rho_A \rho_B), \label{eq:overlap}
\end{align}
where $\rho_A$ and $\rho_B$ denote the output states of two quantum registers subjected to the same circuit instance but independent realizations of noise. This quantity provides a basis-independent measure of similarity between quantum states. In the special case where $\rho_A = \rho_B$, the overlap reduces to the purity $\mathrm{Tr}(\rho^2)$, while if one state is taken to be a pure ideal state, the overlap is equal to the average state fidelity between $\rho_A$ and $\rho_B$ \cite{Jozsa_1994}.

Overlap observables arise naturally in a variety of contexts, including purity estimation and state comparison, and directly quantify the similarity between two quantum states through a joint measurement \cite{Ekert_2002, Buhrman_2001}. Importantly, they can be estimated directly from experimental measurements without requiring classical computation of ideal output distributions \cite{destructive_swap, chen2023unitarity}. This removes the principal scalability bottleneck associated with XEB while retaining the ability to probe the behavior of general random circuit ensembles.

In this work, we leverage overlap as a benchmarking primitive, replacing the ideal-probability comparisons used in \ac{xeb} with a directly measurable quantity. This enables characterization of system performance using general circuit ensembles without relying on classical simulation. The specific measurement circuits used to estimate this quantity are described in Sec.~\ref{sec:protocol}.

\section{Related Work}

\begin{figure*}[hbt!]
     \centering
     \begin{subfigure}[b]{0.25\textwidth}
         \centering
         \includegraphics[width=3.5cm, height=2.53cm]{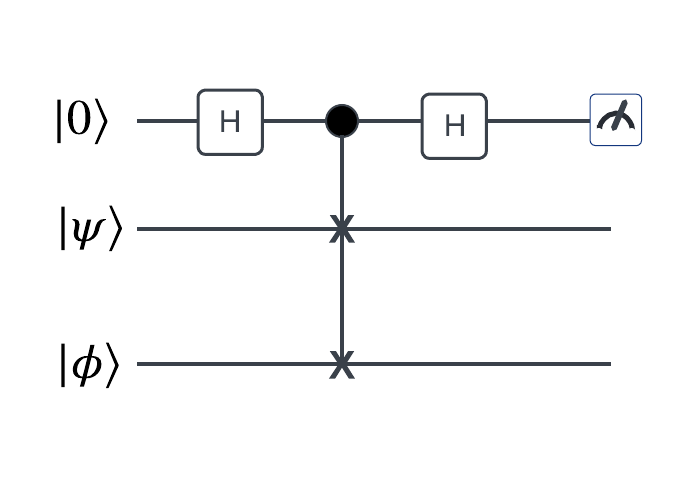}
         \caption{Swap test}
         \label{fig:swap_test_circuit}
     \end{subfigure}
     \begin{subfigure}[b]{0.25\textwidth}
         \centering
         \includegraphics[width=3.5cm, height=3.28cm]{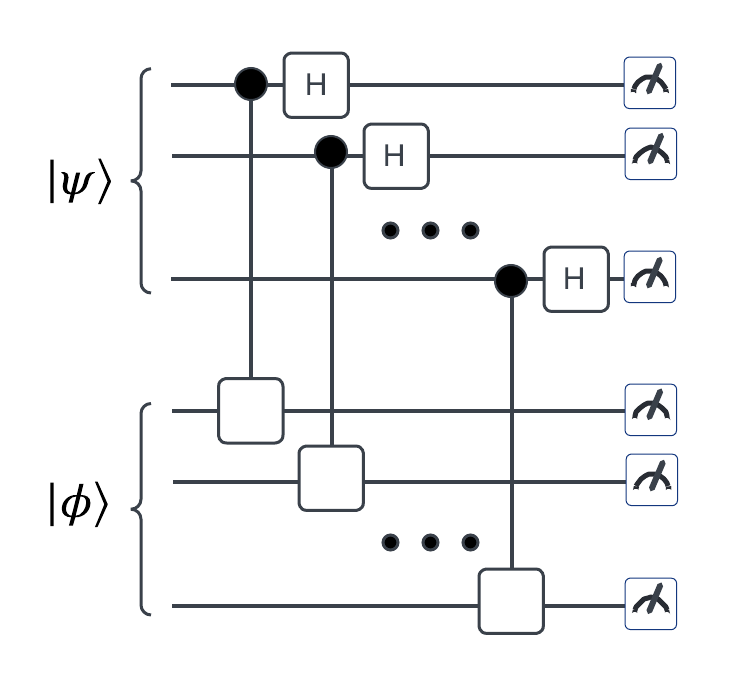}
         \caption{Bell-basis measurement}
         \label{fig:bell_pair_circuit}
     \end{subfigure}        
    \begin{subfigure}[b]{0.4\textwidth}
         \centering
         \includegraphics[width=7cm, height=3.12cm]{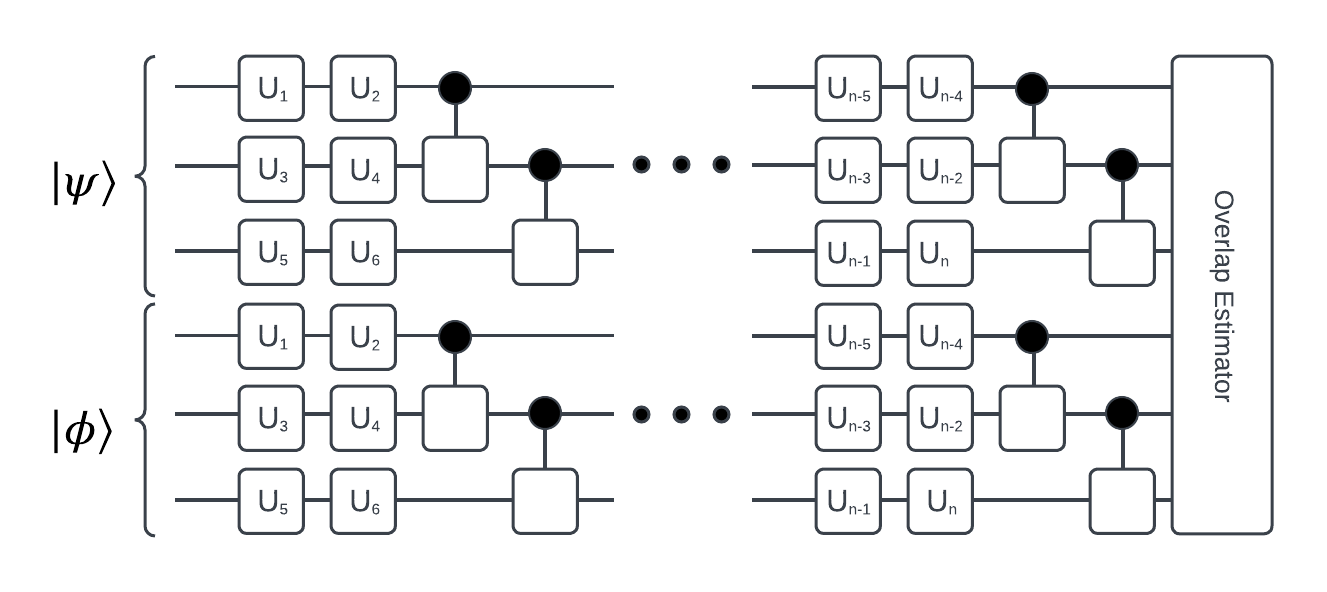}
         \caption{EPCL circuit construction}
         \label{fig:epcl_circuit}
     \end{subfigure}
        \caption[EPCL circuit construction and overlap measurement implementations.]
        {Overlap measurement and random circuit construction used in \ac{epcl}.
        (a) The swap test estimates the overlap $\mathrm{Tr}(\rho_A\rho_B)$ using an ancilla and a controlled-SWAP operation between the two registers.
        (b) The pairwise Bell-basis measurement replaces the controlled-SWAP with local Bell-basis measurements between corresponding qubits; classical post-processing of the measurement outcomes recovers the state overlap.
        (c) \Ac{epcl} applies the same random circuit instance to two disjoint $n$-qubit registers and estimates the overlap between their noisy output states. The illustrative circuit uses $T=2$ for compactness; the numerical and experimental studies in this work use $T=4$. Independently sampled single-qubit Haar-random gates are interleaved with fixed two-qubit entangling layers, after which either overlap-measurement implementation may be applied.}
        \label{fig:overlap_circuits}
\end{figure*}

\Ac{epcl} sits at the intersection of overlap-based characterization methods and circuit-level benchmarking protocols. We briefly review related work and clarify how our protocol differs in objective and implementation.

Unitarity benchmarking uses second-order information to estimate the unitarity of a noisy channel, a diagnostic quantity that separates coherent and incoherent contributions to the noise \cite{Wallman_2015}. Like \ac{epcl}, this connects naturally to purity- and overlap-derived observables. However, the objective is different. Unitarity benchmarking is designed to diagnose the structure of a noise channel, whereas \ac{epcl} compresses the accumulated effect of noise during circuit execution into a single effective layer-polarization parameter. Thus, \ac{epcl} is not intended to decompose noise into coherent and incoherent components; it provides an operational circuit-level measure of register performance.

The implementation also differs. Unitarity benchmarking obtains second-order information through randomized single-register experiments and classical post-processing. By contrast, \ac{epcl} directly implements an overlap observable using a two-register construction, realized through either a controlled-SWAP circuit or destructive Bell-basis measurements. While two-copy approaches have been discussed as a route to accessing such observables \cite{Wallman_2015}, their use as a practical circuit-level benchmarking primitive introduces additional considerations, including measurement-block error, register covariance, and sensitivity to correlated error processes. These effects are central to the analysis in this work.

Overlap- and purity-based observables have also been studied in quantum learning and channel characterization. For example, \cite{chen2023unitarity} analyzes how second-order quantities accessible through two-copy measurements, such as SWAP tests, can be used to infer properties of quantum channels and characterize sample complexity. In contrast to these learning-oriented settings, \ac{epcl} uses the overlap observable as the terminal measurement in a depth-dependent benchmarking protocol. The goal is not to reconstruct or learn detailed channel properties, but to extract a scalar decay parameter that summarizes circuit-layer performance for a chosen random circuit ensemble.

Among circuit-level benchmarks, \ac{eplg} is especially relevant because it also targets performance at the scale of multi-qubit layers \cite{mckay2023benchmarking}. \Ac{eplg} partitions a connected set of qubits into disjoint layers of two-qubit gates and runs simultaneous direct \ac{rb} \cite{gambetta2012characterization} experiments on those layers . The resulting local process fidelities are combined multiplicatively to form a layer fidelity, which is then normalized to obtain an error-per-layered-gate value. This construction has important practical advantages: it is scalable, inherits many of the robustness properties of \ac{rb}, provides pair-resolved diagnostic information, and can reveal some crosstalk effects through simultaneous execution.

The distinction between \ac{eplg} and \ac{epcl} is therefore not simply that one is \ac{rb}-based and the other is overlap-based. The two protocols aggregate information in different orders. \Ac{eplg} first estimates local decay parameters on disjoint sublayers and then reconstructs a layer-level metric from those components. \Ac{epcl}, in contrast, first applies identical random circuits to two copies of the full benchmark register and then estimates a single decay parameter from the resulting overlap decay. Consequently, \ac{epcl} does not provide the local pair-level diagnostic information available from \ac{eplg}. Its advantage is instead that the reported scalar is obtained directly from a full-register circuit ensemble, without first decomposing the benchmark into independent local \ac{rb} experiments. This makes \ac{epcl} a complementary circuit-level probe, particularly for studying aggregate register performance and error mechanisms whose effects may not be fully captured by a product of local layer estimates.

\section{EPCL Protocol} \label{sec:protocol}

\Ac{epcl} combines ideas from XEB, purity estimation, and randomized benchmarking to estimate the average polarization of an $n$-qubit circuit layer. In this section we describe the EPCL protocol in detail. We first present the core benchmarking algorithm, then describe the overlap measurements used to estimate the observable, and finally specify the random circuit ensemble used in the experiment. 

\subsection{Algorithm}\label{sec:algorithm}

The \ac{epcl} protocol estimates the polarization of an $n$-qubit circuit layer by measuring the overlap between two quantum registers that independently undergo identical random circuit evolutions. Throughout this work, $n$ denotes the size of a single benchmark register. Because \ac{epcl} operates on two copies of the benchmarked system, the total benchmark size is
\begin{align}
    N = 2n.
\end{align}
Unless otherwise stated, references to benchmark size refer to $N$, while the theoretical analysis is expressed in terms of the single-register size $n$.

The central algorithm proceeds as follows:

\begin{enumerate}
    \item Select two disjoint $n$-qubit registers from the quantum processor, denoted $A$ and $B$.

    \item Select a set of circuit depths $d = (1,\ldots,d_{\max})$.

    \item For each depth $d_i$, generate $L$ random circuit sequences 
    $\{S_{j}\}_{j=1}^{L}$ acting on $n$ qubits, drawing gates from a universal gate set leading to circuit layers $V$,
    \begin{align}
        S_j = V_{d_i-1} \circ \ldots \circ V_0 .
    \end{align}

    \item For each sequence $S_j$, apply the same circuit instance to both registers ideally by implementing a unitary $U_{d,j}$ where $d$ is the circuit depth. After the circuits are executed, estimate the overlap between the two output states,
    \begin{align}
        S(U_{d,j}) = \mathrm{Tr}(\rho_A(U_{d,j})\rho_B(U_{d,j})),
    \end{align}
    where $\rho_A(U)$ and $\rho_B(U)$ denote the noisy output states on registers $A$ and $B$.

    \item Average the overlap estimates over the $L$ random circuits for each depth, producing an estimate of
    \begin{align}
        \mathbb{E}_{U_d}[S(U_{d})].
    \end{align}

    \item Fit the resulting depth-dependent decay to the model
    \begin{align}
        \mathbb{E}_{U_d}[S(U_d)] =
        \frac{1}{2^{n}}
        +
        \left(A^2-\frac{1}{2^{n}}\right)p^{2d},
    \end{align}
    where $p$ is the effective process polarization and $A$ captures
    depth-independent state-preparation and measurement contributions to the signal amplitude.
\end{enumerate}

\Ac{epcl} does not require classical simulation of the ideal output distribution. Instead, it relies on direct estimation of the state overlap between two identically prepared circuit outputs.

\subsection{Overlap measurement}

Central to the \ac{epcl} protocol is estimating the state overlap defined in Eq.~\ref{eq:overlap}. In this work we consider two measurement circuits that allow this quantity to be estimated: the swap test and a circuit based on pairwise Bell-basis measurements. The physical qubit footprint of \ac{epcl} depends on the overlap measurement implementation. The Bell-basis measurement operates directly on the $N=2n$ benchmark qubits, while the swap-test implementation requires one additional ancilla qubit, for a total physical footprint of $N+1$ qubits.

\subsubsection{Swap test}

The swap test \cite{Buhrman_2001}, depicted in Fig.~\ref{fig:swap_test_circuit}, uses an ancilla qubit to probe the overlap between two states $\ket{\psi}$ and $\ket{\phi}$. After preparing the ancilla in the $\ket{+}$ state, a controlled-SWAP operation is applied between the two registers before measuring the ancilla in the computational basis. The probability of measuring the ancilla in the state $\ket{0}$ is
\begin{align}
    P(0) = \frac{1}{2} + \frac{1}{2}|\braket{\psi|\phi}|^2 .
\end{align}

For mixed states $\rho$ and $\sigma$, the same circuit yields
\begin{align}
    P(0) = \frac{1}{2} + \frac{1}{2}\mathrm{Tr}(\rho\sigma),
\end{align}
allowing the overlap to be obtained from measurements of the ancilla. While the swap test requires measuring only a single qubit, the controlled-SWAP operation acting on two $n$-qubit registers generally requires $\mathcal{O}(n)$ two-qubit gates.

\subsubsection{Pairwise Bell measurements}

An alternative implementation replaces the controlled-SWAP with pairwise Bell-basis measurements across corresponding qubits of the two registers \cite{destructive_swap}, as shown in Fig.~\ref{fig:bell_pair_circuit}. In this circuit each pair of qubits is rotated into the Bell basis and measured. The measurement outcomes are classically post-processed to reconstruct a scalar estimator equivalent to
\begin{align}
    \frac{1}{2} + \frac{1}{2}\mathrm{Tr}(\rho\sigma).
\end{align}

This approach avoids the controlled-SWAP operation and uses only local two-qubit gates, at the cost of increased measurement complexity since all $2n$ qubits must be measured and additional shots may be required to estimate the overlap with comparable statistical precision.

\subsection{Random circuit construction}

There are many possible constructions of random circuits suitable for
benchmarking protocols \cite{Weinstein_2008, Boixo2018}. In this work we use a layered random circuit architecture consisting of alternating single-qubit and two-qubit entangling layers. 

In each cycle, single-qubit gates are drawn independently from the Haar measure on $\mathrm{SU}(2)$ and applied to every qubit in the register. These are followed by a fixed sequence of two-qubit entangling gates that couple neighboring qubits across the register. The same circuit instance is applied independently to both registers in the \ac{epcl} protocol as illustrated in Fig.~\ref{fig:epcl_circuit}. 

This construction provides a simple random-circuit ensemble that is straightforward to implement on hardware. Layered pseudo-random circuits of this form can approach Haar-like second moment statistics, and with sufficient mixing, scramble local noise towards an effective global white-noise description \cite{Weinstein_2008, Dalzell_2024}, motivating the ensemble-averaged depolarizing model adopted in Sec.~\ref{sec:theory}. 

Following~\cite{Boixo2018}, we introduce a parameter $T$ that sets the number of single-qubit random layers between applications of the entangling layer. Tuning $T$ provides control over the rate at which errors mix under the circuit ensemble. In this work we use $T=4$. While the theoretical interpretation of the protocol is insensitive to this choice, the value of $T$ affects circuit depth and may influence performance on specific hardware platforms.

\section{EPCL Theory} \label{sec:theory}

In this section we establish the theoretical foundation of \ac{epcl}. Building on prior work in random circuit benchmarking, we adopt a depolarizing-channel model for ensemble-averaged noisy random circuits and show that it can be consistently applied within the \ac{epcl} protocol, despite the use of two registers and an overlap observable. We then show that \ac{epcl} probes the corresponding polarization via an estimator of the state overlap that is robust to residual Pauli structure and \ac{spam} errors. 

\subsection{Ensemble averaged noise model for \ac{epcl}} \label{sec:theory_ensemble_avg}

In \ac{epcl} we adopt the ensemble-averaged depolarizing model established in the theory of \ac{xeb} \cite{Arute2019}. In that work it is argued and numerically demonstrated that averaging over an ensemble of sufficiently random noisy circuits suppresses Pauli anisotropy, leading to an effective depolarizing description for the relevant observable. We do not re-derive their result, but rather briefly summarize and restate it before discussing its applicability to \ac{epcl}. 

For a fixed random circuit instance $U$ with $m$ layers, the noisy evolution generally induces a Pauli channel with dense support. However, as shown in \cite{Arute2019}, averaging over the circuit ensemble after undoing the ideal evolution suppresses all non-identity Pauli components, leaving only the identity, which is invariant under conjugation. The ensemble-averaged evolution may therefore be written as
\begin{align}
    \mathbb{E}_U[U^{\dagger} \rho_f U] = p_m \rho_0 + (1-p_m)\frac{I}{2^n} \label{eq:xeb_dep_model}
\end{align}
where $\rho_0$ is the initial state, and $\rho_f$ is the state after the $m$-layer random circuit. Inspection of Eq.~\ref{eq:xeb_dep_model} shows it corresponds to an effective depolarizing channel with polarization $p_m$. 

Unlike \ac{xeb}, \ac{epcl} does not average single-register measurement outcomes. Instead, it estimates the overlap between two quantum registers subjected to the same circuit instance. Assuming that the two registers are non-interacting and experience independent noise realizations conditioned on the applied circuit, measurement of the overlap yields an estimator whose expectation equals $\mathrm{Tr}(\rho_A(U)\rho_B(U))$ in the absence of measurement error. We explore deviations from these assumptions and analyze their impact on the estimator in Sec.~\ref{sec:robustness_and_error_conditions}.

The question then is whether the joint overlap measurement invalidates the depolarizing description established in \cite{Arute2019}.

Let $X(U)$ denote the overlap estimator random variable associated with a fixed circuit instance $U$. The expected overlap for a fixed $U$ is then
\begin{align}
    \mathbb{E}[X(U)] = \mathrm{Tr}(\rho_A(U)\rho_B(U)),
\end{align}
where $\rho_A(U) = \mathcal{E}_A (U \rho_0 U^{\dagger})$ and $\rho_B(U) = \mathcal{E}_B (U \rho_0 U^{\dagger})$ are the noisy output states on registers $A$ and $B$, respectively. 

Averaging over an ensemble of random circuits therefore probes
\begin{align}
    \mathbb{E}_U[X] 
    = \mathbb{E}_U[\mathrm{Tr}(\rho_A(U)\rho_B(U))].
    \label{eq:epcl_probe}
\end{align}

Next, we follow the idea established in \cite{Arute2019} and uncompute the noisy evolution. Let $\hat{\rho}_A(U) = U^{\dagger} \rho_A(U) U$ and $\hat{\rho}_B(U) = U^{\dagger} \rho_B(U) U$ be the uncomputed evolution. This uncompute step does not change the ensemble average due to the cyclicity of the trace and so 
\begin{align}
    \mathbb{E}_U[X] 
    = \mathbb{E}_U[\mathrm{Tr}(\hat{\rho}_A(U)\hat{\rho}_B(U))].
    \label{eq:epcl_probe}
\end{align}

This quantity can be decomposed as
\begin{align}
    \mathbb{E}_U[X]
    &= \mathrm{Tr}(\bar{\rho}_A \bar{\rho}_B)
    + \mathbb{E}_U\!\left[
        \mathrm{Tr}\!\left(
            (\hat{\rho}_A(U) - \bar{\rho}_A)
            (\hat{\rho}_B(U) - \bar{\rho}_B)
        \right)
      \right],
    \label{eq:epcl_expect}
\end{align}
where $\bar{\rho}_A \equiv \mathbb{E}_U[\hat{\rho}_A(U)]$ and 
$\bar{\rho}_B \equiv \mathbb{E}_U[\hat{\rho}_B(U)]$.

The second term in Eq.~\ref{eq:epcl_expect} quantifies the deviation between the ensemble-averaged overlap and the overlap of the ensemble-averaged states. It captures correlations between register fluctuations across the circuit ensemble and represents the correction introduced by \ac{epcl}'s joint two-register measurement relative to the single-register observables considered in \ac{xeb}. We define this term as the register covariance,
\begin{align}
    \Delta_{\mathrm{reg}}
    \equiv
    \mathbb{E}_U\!\left[
        \mathrm{Tr}\!\left(
            (\hat{\rho}_A(U) - \bar{\rho}_A)
            (\hat{\rho}_B(U) - \bar{\rho}_B)
        \right)
    \right].
\end{align}

Under sufficiently mixing random circuits and independent noise realizations across the two registers, the register covariance is expected to be negligible. In this regime, the ensemble-averaged overlap estimate is well approximated by
\begin{align}
    \mathbb{E}_U[X] \approx \textrm{Tr}(\bar{\rho}_A \bar{\rho}_B).\label{eq:epcl_approx}
\end{align}

In the next section, we compute this observable explicitly and show how \ac{epcl} provides an operational estimator of polarization.

\subsection{\Ac{epcl} observable under the depolarizing-channel model}

Let the generalized fixed circuit instance $U$ described in the preceding section now be a fixed $m$-layer random circuit, and let $D=2^n$ denote the Hilbert-space dimension of a single $n$-qubit benchmark register. 

Recall that each term in the trace in Eq.~\ref{eq:epcl_approx} is the expectation over random circuits of the uncomputed ideal evolution. Under the depolarizing-channel model used in Eq.~\ref{eq:xeb_dep_model}, these terms can be written as
\begin{align}
    \bar{\rho}_A = p_A^m \rho_0 + (1-p_A^m)\frac{I}{D} \\
    \bar{\rho}_B = p_B^m \rho_0 + (1-p_B^m)\frac{I}{D} \\
\end{align}
where $p_A$ and $p_B$ are the effective layer polarizations for register $A$ and $B$ respectively. 

Inserting these definitions into Eq.~\ref{eq:epcl_approx} the ensemble averaged overlap estimate can be expanded to

\begin{align}
    \mathrm{Tr}(\bar{\rho}_A \bar{\rho}_B)
    = \frac{1}{D}
    + (p_A p_B)^m
    \left(A^2-\frac{1}{D}\right)
    \label{eq:epcl_fit_func}
\end{align}
where $\mathrm{Tr}(I)=D$ and $A^2=\mathrm{Tr}(\rho_0^2)$. In the absence of state-preparation errors, $\rho_0$ is pure and therefore $A=1$.

Thus, the ensemble-averaged \ac{epcl} overlap estimator directly probes the product $p_Ap_B$ of the two register polarizations. We therefore report $p_{\mathrm{eff}} = \sqrt{p_Ap_B}$, which may be interpreted as the effective $n$-qubit register polarization and reduces to $p$ in the symmetric case $p_A=p_B$.

Under the same depolarizing-channel interpretation, the fitted effective polarization can also be converted into fidelity- and error-based quantities.  If one circuit layer is modeled by the effective depolarizing channel
\begin{align}
\mathcal{D}_{p_{\mathrm{eff}}}(\rho)
=
p_{\mathrm{eff}}\rho
+
(1-p_{\mathrm{eff}})\frac{I}{D},
\end{align}
then $p_{\mathrm{eff}}$ is the process polarization of the effective layer. The corresponding entanglement fidelity, often equivalently referred to as process fidelity, is
\begin{align}
F_{\mathrm{e,layer}}
=
\frac{1+(D^2-1)p_{\mathrm{eff}}}{D^2}.
\end{align}
Using the standard relation between entanglement fidelity and average gate fidelity \cite{Nielsen_2002}, this gives the effective average layer fidelity
\begin{align}
F_{\mathrm{avg,layer}}
&=
\frac{D F_{\mathrm{e,layer}}+1}{D+1} \\
&=
\frac{1+(D-1)p_{\mathrm{eff}}}{D}.
\end{align}
Equivalently, the effective average layer infidelity is
\begin{align}
r_{\mathrm{avg,layer}}
=
1-F_{\mathrm{avg,layer}}
=
\frac{D-1}{D}(1-p_{\mathrm{eff}}).
\end{align}
Throughout this work we report $p_{\mathrm{eff}}$ as the primary \ac{epcl} estimate because it is the decay parameter directly extracted from the overlap signal. The derived quantities $F_{\mathrm{e,layer}}$, $F_{\mathrm{avg,layer}}$, and $r_{\mathrm{avg,layer}}$ provide process-fidelity, average-fidelity, and error-based interpretations when comparison to conventional benchmark metrics is useful.

\section{Robustness and error conditions}
\label{sec:robustness_and_error_conditions}
The accuracy of \ac{epcl} as an estimator of the effective $n$-qubit polarization relies on several assumptions, including ensemble-induced depolarization at the level of the \ac{epcl} observable, approximately Markovian noise, conditional independence of the two registers, and a negligible contribution from the register covariance term identified in Sec.~\ref{sec:theory_ensemble_avg}. The assumptions of ensemble-induced depolarization and approximately Markovian noise are closely related to those commonly employed in analyses of \ac{xeb}, \ac{rb}, and other random-circuit benchmarks \cite{Boixo2018,Magesan_2011,Dalzell_2024}. Conditional independence and negligible register covariance arise specifically from the two-register construction used by \ac{epcl} and are examined below. Violations of these assumptions can introduce deviations from the idealized \ac{epcl} model and affect estimation of the effective polarization. In this section, we analyze the principal sources of such deviations.

\subsection{Finite sampling error}

Both finite sampling of the random circuit ensemble and the per-random circuit observable introduce statistical uncertainty. For $k$ independently sampled random circuit instances with $N_{\mathrm{shots}}$ shots per instance, the variance of the ensemble-averaged estimator can be decomposed into contributions from circuit-to-circuit variation and shot noise,
\begin{align}
\operatorname{Var}(\hat{S})
=
\frac{\sigma_{\mathrm{circ}}^2}{k}
+
\frac{\sigma_{\mathrm{shot}}^2}{kN_{\mathrm{shots}}},
\end{align}
where $\hat{S}$ is the estimated ensemble-averaged observable, $\sigma_{\mathrm{circ}}^2$ denotes the variance of the expected observable across circuit instances, and $\sigma_{\mathrm{shot}}^2$ denotes the ensemble-averaged single-shot measurement variance. When circuit-to-circuit variation is negligible, as expected under the idealized ensemble-averaged model, the statistical uncertainty approaches $O(1/\sqrt{kN_{\mathrm{shots}}})$. Otherwise, increasing the number of shots per circuit cannot eliminate uncertainty arising from finite sampling of the circuit ensemble.

\subsection{Insufficient circuit scrambling}
Pauli anisotropy represents a second potential source of model mismatch. If the pseudo-random circuit ensemble does not sufficiently scramble the Pauli frame, ensemble averaging fails to induce a uniform effective twirl, leaving residual Pauli-frame bias in the dressed noise channel. The EPCL protocol employs a brickwork circuit architecture composed of layers of single-qubit Haar-random gates interleaved with fixed two-qubit entangling gates. This class of circuits has been extensively studied and is known to rapidly randomize Pauli operators and suppress structured error components under ensemble averaging \cite{Arute2019, Boixo2018, Morvan2024, haferkamp2022random, Harrow2023, Schuster2025}. While incomplete mixing can in principle lead to residual anisotropy, we expect such effects to be minimal for the circuit constructions considered here.

\subsection{Readout and measurement error}

Readout and measurement-circuit errors affect \ac{epcl} by perturbing the measured overlap observable. Readout errors act after the quantum circuit and distort the observed bit-string distribution, while measurement-circuit errors arise from imperfect gates used to implement the overlap measurement itself. Both mechanisms can bias the fitted effective polarization, but their effects need not correspond to a unique or physically identifiable modification of the decay model.

In simple cases, depth-independent measurement imperfections may be approximated as an affine distortion of the measured overlap, equivalent to changing the contrast and offset of the fitted exponential. In this sense, the fit parameter $A$ in Eq.~\ref{eq:epcl_fit_func} can absorb measurement-induced bias and may serve as a useful proxy for changes in aggregate \ac{spam} behavior. We do not interpret $A$ as a standalone \ac{spam} metric, nor does it identify the mechanisms responsible for the distortion. However, comparisons of $A$ across otherwise similar \ac{epcl} benchmarks, such as different registers or repeated measurements of the same register over time, can reveal changes in the combined state-preparation, readout, and measurement-circuit contribution to the observed overlap.

In general, however, additional fit parameters that modify the contrast or offset of the decay would aggregate multiple error mechanisms and would not be independently identifiable from \ac{epcl} decay data alone. We therefore do not attempt to separate readout, state-preparation, and measurement-circuit contributions at the level of the fitted decay model. Instead, we quantify the impact of readout and measurement-circuit errors directly through numerical simulation in Sec.~\ref{sec:measurement-error}. Readout errors can also be reduced using established measurement-error mitigation techniques \cite{Bravyi_2021, Temme_2017, Endo_2018, ball2020}. In Sec.~\ref{sec:hardware}, we apply measurement-error mitigation to hardware data and evaluate its effectiveness in reducing bias in \ac{epcl}.

\subsection{Register covariance}

The derivation of the \ac{epcl} estimator assumes that the two registers evolve independently so that the register covariance term introduced in Eq.~\ref{eq:epcl_expect} is negligible. In experimental systems this assumption may be violated if correlated error processes couple the two registers.

One important mechanism capable of generating such correlations is inter-register crosstalk, where coherent interactions couple qubits belonging to different registers during circuit execution. In this regime the factorized polarization model used to interpret the \ac{epcl} observable may no longer strictly apply. The practical impact of such correlations on the \ac{epcl} estimator is investigated numerically in Sec.~\ref{sec:simulation}.

\section{Numerical simulations} \label{sec:simulation}

We evaluate the performance of \ac{epcl} through numerical simulations of noisy quantum circuits. The analysis is organized by noise model, enabling a systematic study of the estimator under depolarizing, coherent, and crosstalk-dominated error processes. For each setting, we assess the validity of the underlying modeling assumptions and quantify the resulting bias in the extracted polarization. We further examine the impact of measurement errors and the robustness of the overlap estimator to \ac{spam}-like imperfections.

\subsection{Methodology and metrics}

This section defines the simulation framework and evaluation metrics used throughout the numerical analysis.

\subsubsection{Simulation setup.}
Numerical simulations are performed using Qiskit Aer \cite{qiskit} on a $9$-qubit ring topology. Unless otherwise stated, overlap estimation is implemented using the swap-test measurement circuit. The system is partitioned into two disjoint benchmark registers of size $n=4$, with the remaining qubit serving as the ancilla required for the controlled-SWAP operation. Each register is initialized in the state $\ket{0}^{\otimes 4}$.

Random circuits are constructed in discrete layers. Each layer consists of $T=4$ independently sampled single-qubit Haar-random gates applied to every qubit in the register, followed by a ladder of two-qubit $CX$ gates. Circuit depth is defined by the number of such layers, consistent with the theoretical model. Unless otherwise stated, simulations are performed at depths $m\in[0,1,2,4,8,16,32]$.

At each depth, overlap estimation is averaged over $20$ independently sampled random circuits, with $2048$ measurement shots per circuit. \ac{spam} errors are excluded by default to isolate gate-level noise processes. The impact of measurement-circuit and readout errors is examined separately in Sec.~\ref{sec:measurement-error}.

\begin{figure*}[hbt!]
     \centering
     \begin{subfigure}[b]{0.32\textwidth}
         \centering
         \includegraphics[width=\textwidth]{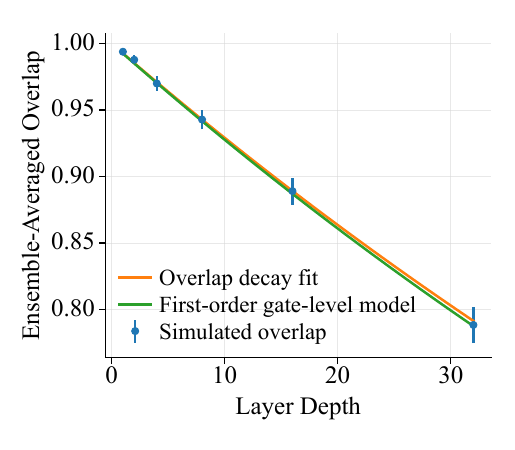}
         \caption{Weak-noise regime}
         \label{fig:local_depol_1e4_fit}
     \end{subfigure}
     \begin{subfigure}[b]{0.32\textwidth}
         \centering
         \includegraphics[width=\textwidth]{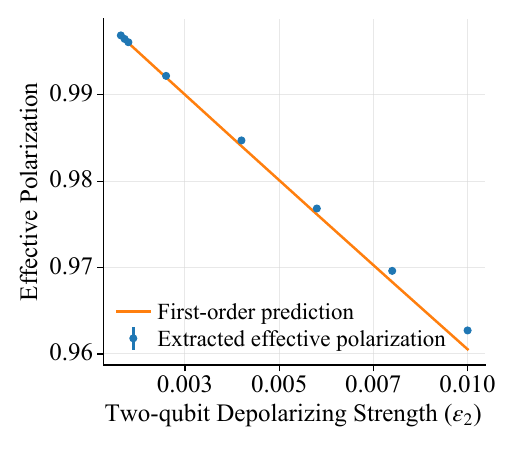}
         \caption{Noise-strength sweep}
         \label{fig:local_depol_sweep}
     \end{subfigure}
     \begin{subfigure}[b]{0.32\textwidth}
         \centering
         \includegraphics[width=\textwidth]{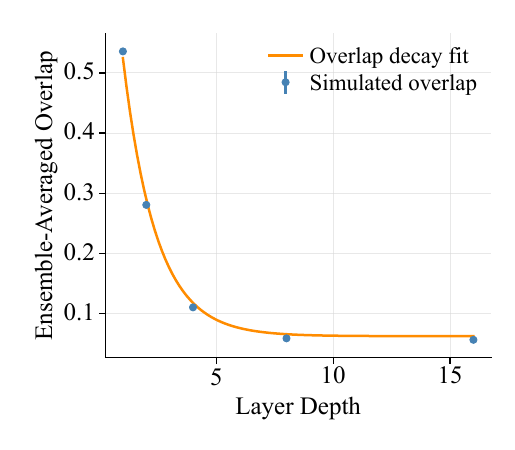}
         \caption{Strong-noise regime}
         \label{fig:local_depol_1e1_fit}
     \end{subfigure}
        \caption[EPCL under local depolarizing noise.]
        {\Ac{epcl} under local depolarizing noise, demonstrating agreement with the first-order baseline model in the weak-noise regime and stable decay behavior beyond the validity of that approximation.
        (a) For $\varepsilon_1 = 10^{-4}$ and $\varepsilon_2 = 10^{-3}$, the \ac{epcl} fit closely agrees with the simulated overlap decay and yields $\widehat{p}_{\mathrm{eff}}=0.99601$, compared with the first-order prediction $p_{\mathrm{eff}}^{\mathrm{FO}}=0.99599$.
        (b) Sweeping the two-qubit depolarizing strength while maintaining $\varepsilon_1=0.1\varepsilon_2$ produces close agreement at low noise strengths, with systematic deviation emerging near $\varepsilon_2 \approx 6\times10^{-3}$ as higher-order Pauli fault contributions become significant and the first-order approximation breaks down.
        (c) At the intentionally large noise strength $\varepsilon_2=10^{-1}$, the overlap remains well described by the exponential decay model and approaches the maximally mixed-state limit, demonstrating stable extraction of an effective polarization well beyond the weak-noise regime.}
        \label{fig:local_depol_model}
\end{figure*}

\subsubsection{Evaluation criteria}
The numerical evaluation of \ac{epcl} is divided into two regimes based on the expected validity of a first-order gate-level baseline approximation.

In the weak-noise regime, where higher-order fault combinations are expected to be negligible, \ac{epcl} is evaluated by comparison against this baseline model. Exact analytical prediction of effective polarization in noisy random circuits generally requires tracking the propagation and interaction of Pauli fault configurations, which rapidly becomes intractable as circuit size and noise strength increase. We therefore use a first-order approximation that captures the leading contribution from stochastic Pauli noise.

For a gate-level Pauli channel of the form
\begin{align}
    \rho(g) \approx p_g \rho_{\mathrm{id}}(g) + (1-p_g)\chi,
\end{align}
where $p_g$ denotes the identity Pauli weight associated with gate $g$, the effective polarization of a circuit layer is approximated as
\begin{align}
    p_{\mathrm{layer}} = \prod\limits_{g \in \mathrm{layer}} p_g.
\end{align}
Circuit-level polarization is then obtained through multiplicative composition across layers.

This approximation neglects higher-order contributions to the identity weight arising from multiple Pauli faults within or across layers. These effects become increasingly significant as circuit size and noise strength increase. The expected number of non-identity faults across an $m$-layer circuit is approximately
\begin{align}
    m\big(nT(1-p_1) + C(1-p_2)\big) \ll 1,
\end{align}
which therefore defines the regime in which quantitative agreement with the first-order model is expected. Here, $nT$ is the number of single-qubit gates per layer, $C$ is the number of two-qubit gates per layer, and $p_1$ ($p_2$) denotes the single- (two-) qubit identity Pauli weight. 

Within this regime, agreement between the extracted \ac{epcl} effective polarization and the first-order model provides a consistency check: the fitted decay agrees with the leading-order prediction in the parameter regime where higher-order corrections are expected to be small.

Outside the weak-noise regime, the first-order model is no longer expected to provide an accurate reference. In these cases, \ac{epcl} is instead evaluated through the stability and physical interpretability of the extracted decay behavior. Specifically, we assess whether the overlap observable remains well-described by a single-parameter exponential decay and whether the extracted trajectories converge toward the expected mixed-state limit as noise strength increases.

Throughout the numerical simulations, we distinguish between the effective polarization extracted from simulated \ac{epcl} data and the polarization predicted by the first-order gate-level baseline model. We denote the fitted \ac{epcl} estimate by
\begin{align}
    \widehat{p}_{\mathrm{eff}},
\end{align}
and the first-order baseline prediction by
\begin{align}
    p_{\mathrm{eff}}^{\mathrm{FO}}.
\end{align}
The former is obtained by fitting the depth-dependent overlap signal to Eq.~\ref{eq:epcl_fit_func}, whereas the latter is computed directly from the injected gate-level noise parameters using the first-order composition rule above. 

\begin{figure*}[hbt!]
     \centering
     
         \includegraphics[width=0.45\textwidth]{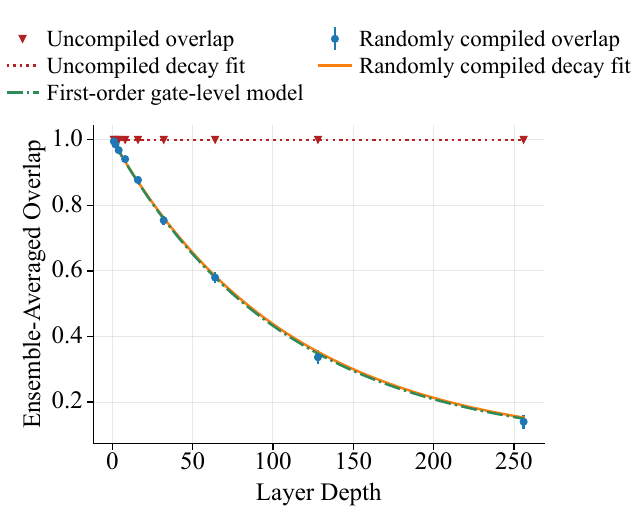}
     
        \caption[EPCL sensitivity to coherent errors with and without Pauli twirling.]
        {Effect of Pauli twirling on \ac{epcl} sensitivity to a fixed coherent error in the entangling layer. A coherent over-rotation of $\theta=\pi/40$ is applied to the target qubit of each $CX$ gate. Without Pauli twirling, the overlap exhibits essentially no depth-dependent decay, indicating that the unmodified \ac{epcl} protocol is insensitive to this coherent error structure. With Pauli twirling, the coherent error is randomized into an effective stochastic channel, restoring the exponential decay used by \ac{epcl} to estimate effective polarization. The extracted polarization, $\widehat{p}_{\mathrm{eff}}=0.9955$, closely agrees with the first-order prediction $p_{\mathrm{eff}}^{\mathrm{FO}}=0.9954$.}
        \label{fig:coherent_error}
\end{figure*}

\subsubsection{Statistical uncertainty}
Statistical uncertainty arises from both finite circuit sampling and finite measurement shots. Where reported, error bars are estimated via bootstrap resampling of circuit instances, with shot noise incorporated at the level of individual circuit measurements. Unless otherwise stated, these uncertainties are small relative to the observed systematic effects.

\subsection{Local depolarizing channel} \label{sec:local-depol}

We begin with a local depolarizing noise model, which provides a controlled setting for evaluating \ac{epcl} under the assumptions of the first-order gate-level baseline approximation. This study serves two purposes: validating agreement with the baseline model in the weak-noise regime where higher-order fault contributions are negligible, and assessing the stability of the \ac{epcl} estimator as those assumptions break down.

The simulated noise model is
\begin{align}
    \mathcal{D}_{\varepsilon}(\rho) = (1-\varepsilon)\rho + \varepsilon \frac{I}{D},
\end{align}
where $\varepsilon$ is the depolarizing strength and $D=2^n$. Depolarizing noise is applied at the gate level, with a single-qubit channel following each Haar-random single-qubit gate and a two-qubit channel following each entangling gate. The single-qubit depolarizing strength $\varepsilon_1$ is chosen to be one order of magnitude smaller than the two-qubit strength $\varepsilon_2$, reflecting the typical hierarchy of gate error rates observed in contemporary quantum hardware.

Consistent with the methodology described above, no depolarizing noise is applied to the overlap measurement circuit in this section in order to isolate gate-level noise processes within the benchmarked circuit.

We first consider a hardware-relevant weak-noise regime with $\varepsilon_1 = 10^{-4}$ and $\varepsilon_2 = 10^{-3}$. Under these conditions, \ac{epcl} yields an extracted effective polarization of $\widehat{p}_{\mathrm{eff}} = 0.99601$, in close agreement with the first-order baseline prediction of $p_{\mathrm{eff}}^{\mathrm{FO}} = 0.99599$. As shown in Fig.~\ref{fig:local_depol_1e4_fit}, both the \ac{epcl} fit and the baseline model closely track the simulated decay, consistent with the expected validity of the first-order approximation in this regime.

We next sweep the depolarizing strength over the range $8 \times 10^{-4} \leq \varepsilon_2 \leq 10^{-2}$ while maintaining the ratio $\varepsilon_1 = 0.1\varepsilon_2$. Figure~\ref{fig:local_depol_sweep} shows strong agreement between \ac{epcl} and the baseline model at low noise strengths, with systematic deviation emerging near $\varepsilon_2 \approx 6 \times 10^{-3}$.

As expected, this deviation reflects breakdown of the first-order approximation rather than failure of the \ac{epcl} estimator. The baseline model neglects higher-order Pauli fault contributions, which become increasingly significant as noise strength increases. These higher-order processes can contribute additional identity weight through partial error cancellation, leading the first-order approximation to systematically underestimate the effective polarization.

Despite the breakdown of the simplified baseline model, the \ac{epcl} estimator remains stable and physically interpretable. The extracted overlap decay remains well-described by a single-parameter exponential model across all simulated depolarizing strengths. This behavior persists even in an intentionally extreme regime with $\varepsilon_2 = 10^{-1}$, shown in Fig.~\ref{fig:local_depol_1e1_fit}, where the fitted decay captures both rapid polarization loss and convergence toward the fully mixed-state limit.

These results demonstrate that \ac{epcl} agrees with the first-order baseline model where that approximation is expected to hold, while continuing to provide stable and interpretable polarization estimates well beyond the weak-noise regime.

\subsection{Coherent channel} \label{sec:coherent-errors}

\begin{figure*}[hbt!]
     \centering
     \begin{subfigure}[b]{0.36\textwidth}
         \centering
         \includegraphics[width=\textwidth]{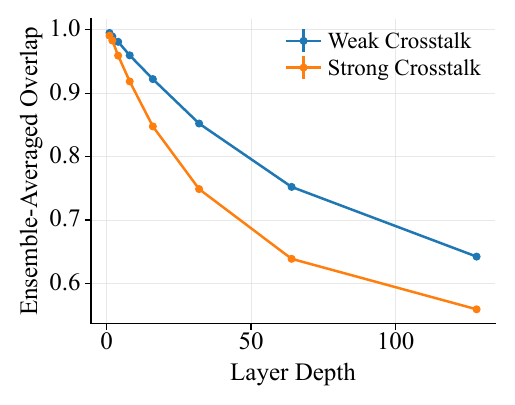}
         \caption{Inter-register overlap decay.}
         \label{fig:inter_register_crosstalk_decay}
     \end{subfigure}
     \begin{subfigure}[b]{0.36\textwidth}
         \centering
         \includegraphics[width=\textwidth]{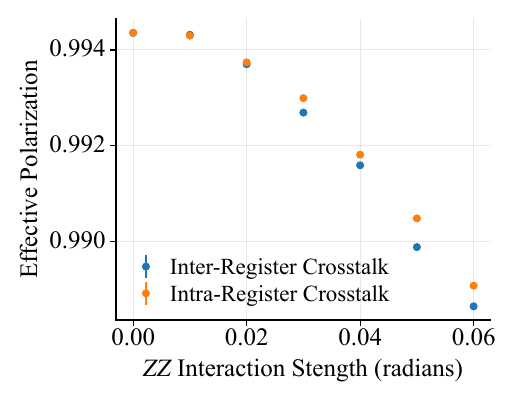}
         \caption{Polarization versus crosstalk strength.}
         \label{fig:inter_and_intra_register_crosstalk_on_peff}
     \end{subfigure}
        \caption[EPCL response to intra- and inter-register crosstalk.]
        {Effect of coherent $ZZ$ crosstalk on \ac{epcl}.
        (a) Ensemble-averaged overlap as a function of circuit depth for inter-register crosstalk at representative weak ($\theta=0.01$) and strong ($\theta=0.06$) interaction strengths. In both cases, the overlap retains a clear exponential depth dependence, allowing an effective polarization to be extracted despite correlations between the two registers.
        (b) Extracted effective polarization as a function of $ZZ$ interaction strength for intra- and inter-register crosstalk. Inter-register coupling produces a systematically lower effective polarization and stronger dependence on interaction strength than intra-register coupling, consistent with contributions from a non-negligible register covariance term that is absent from the independent-register approximation. A fixed two-qubit depolarizing channel is applied after each $CX$ gate in both models, and the overlap measurement circuit is kept error-free to isolate crosstalk during random-circuit execution.}
        \label{fig:local_depol_fits}
\end{figure*}

Depolarizing noise provides a useful stochastic baseline, but real quantum hardware also exhibits coherent error mechanisms arising from systematic over- or under-rotations. Because \ac{epcl} extracts an effective polarization from depth-dependent decay, it is important to understand when coherent errors produce behavior consistent with this stochastic interpretation.

Coherent errors acting on Haar-random single-qubit gates are expected to be effectively randomized under ensemble averaging, yielding a stochastic Pauli-like description consistent with the assumptions underlying \ac{epcl}. This behavior follows from established results in randomized benchmarking and unitary twirling, where coherent errors are transformed into effective stochastic channels under averaging over random unitaries or approximate unitary designs \cite{Arute2019, emerson_2007, magesan_2012, Wallman_2016}.

Coherent errors associated with the fixed two-qubit entangling layers behave differently. Although the surrounding Haar-random single-qubit layers introduce some randomization, this is insufficient to fully wash out the coherent structure of the entangling-layer error. As a result, the induced behavior does not naturally produce the stochastic-like depth-dependent decay that \ac{epcl} uses to infer polarization. We therefore focus our numerical analysis on this more relevant case.

For coherent over-rotation errors associated with two-qubit gates, we model the faulty operation $\tilde{g}$ as
\begin{align}
    \tilde{g}(\rho) = (I \otimes e^{-i\theta X})\;CX \;\rho \;CX^{\dagger}\;(I\otimes e^{i\theta X}),
\end{align}
where $\theta$ denotes the coherent over-rotation magnitude applied to the target qubit. While not fully general, this model captures the key feature of interest: a fixed coherent error associated with the entangling layer.

A straightforward mitigation is to apply Pauli twirling to the coherent error, transforming it into an effective stochastic Pauli channel with an approximately depolarizing structure. This restores the depth-dependent decay behavior to which \ac{epcl} is sensitive.

Figure~\ref{fig:coherent_error} compares \ac{epcl} with and without Pauli twirling for coherent over-rotation errors in the two-qubit entangling layers. A coherent over-rotation of $\theta = \pi/40$ is applied to the target qubit of each $CX$ gate. Using the relationship between average gate fidelity and process polarization~\cite{Hashim_2025}, this corresponds approximately to a depolarizing strength of $\varepsilon \approx 1.6 \times 10^{-3}$, placing the effective stochastic equivalent within the weak-noise regime where the first-order baseline approximation is expected to remain valid.

Without twirling, the overlap observable exhibits essentially no depth-dependent decay, indicating that the unmodified protocol is insensitive to this coherent error structure. With Pauli twirling, the expected exponential decay is restored, and the extracted effective polarization $\widehat{p}_{\mathrm{eff}} = 0.9955$ closely agrees with the first-order prediction $p_{\mathrm{eff}}^{\mathrm{FO}} = 0.9954$.

These results highlight an implementation limitation of the unmodified protocol rather than a fundamental limitation of \ac{epcl}. Coherent errors that do not naturally produce stochastic-like decay can be converted into a form compatible with the protocol through standard twirling techniques, restoring sensitivity and preserving the effective polarization interpretation.

\subsection{Crosstalk noise}

The theoretical treatment in Sec.~\ref{sec:theory_ensemble_avg} showed that the ensemble-averaged \ac{epcl} observable can be approximated by the overlap of the ensemble-averaged register states, yielding the effective polarization model in Eq.~\ref{eq:epcl_approx}. This approximation relies on the assumption that the two registers evolve independently so that register covariance is negligible.

In experimental systems these assumptions may be violated. In this regime correlations between the two registers contribute to the ensemble-averaged overlap, and the approximation must be replaced by the more general expression in Eq.~\ref{eq:epcl_expect}, which includes the register covariance term $\Delta_{\mathrm{reg}}$.

The influence of register covariance on the extracted polarization $p_{\mathrm{eff}}$ depends on the underlying error mechanism. One important source of such correlations is inter-register crosstalk, which can couple qubits belonging to different registers during circuit execution. In this section we investigate how such crosstalk infludes the \ac{epcl} estimator.

We pursue three goals. First, we show that the overlap estimator remains well behaved even when correlations between the two registers are non-negligible. Second, we demonstrate that inter-register errors introduce contributions that are not captured by the approximate model and argue that the resulting deviations are consistent with a non-negligible register covariance term. Finally, we show that \ac{epcl} is sensitive to cross-register correlations in practice, although their presence modifies the operational interpretation of the extracted polarization $p_{\mathrm{eff}}$.

\begin{figure*}[hbt!]
     \centering
     \begin{subfigure}[b]{0.465\textwidth}
         \centering
         \includegraphics[width=\textwidth]{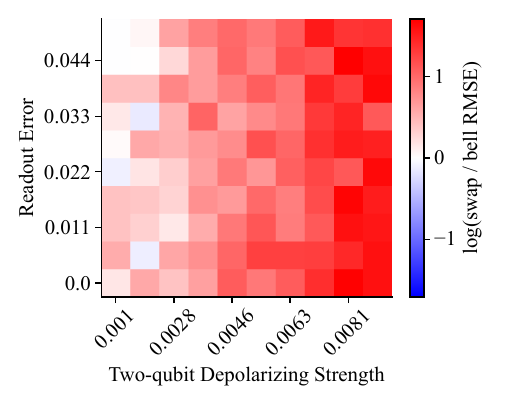}
         \caption{$N=4$}
         \label{fig:measurement_analysis_n5}
     \end{subfigure}
     \begin{subfigure}[b]{0.47\textwidth}
         \centering
         \includegraphics[width=\textwidth]{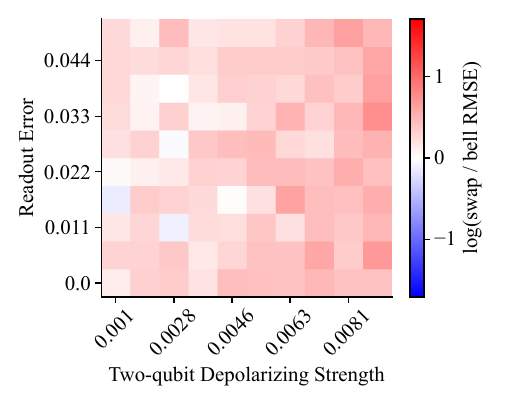}
         \caption{$N=10$}
         \label{fig:measurement_analysis_n11}
     \end{subfigure}
        \caption[Comparison of swap-test and Bell-basis EPCL measurement circuits.]
        {Comparison of the two overlap measurement implementations considered for \ac{epcl} under depolarizing gate noise and readout error. The simulations use physical systems of 5 and 11 qubits, corresponding to $N=4$ and $N=10$ benchmark qubits, respectively. The swap-test implementation uses the remaining qubit as an ancilla, while the Bell-basis implementation does not require the additional qubit. Each point shows $\log(\mathrm{RMSE}_{\mathrm{swap}}/\mathrm{RMSE}_{\mathrm{Bell}})$ for the extracted effective polarization relative to a high-shot ideal reference computed using an error-free measurement circuit; positive values indicate lower estimation error for the Bell-basis implementation.
        (a) For $N=4$, the Bell-basis measurement produces lower \ac{epcl} RMSE across nearly the entire simulated error landscape, with its advantage generally increasing as depolarizing strength increases.
        (b) For $N=10$, the Bell-basis measurement remains favorable across nearly the entire error landscape, although the performance gap narrows as its increased number of measured qubits makes the estimator more sensitive to readout error.}
        \label{fig:measurement_analysis}
\end{figure*}

In simulation we model two types of crosstalk errors. The first model, referred to as \textit{intra-register crosstalk}, injects two local $ZZ$ interactions after each $CX$ gate in the random circuit. One interaction acts between $CX$ target qubit and one of its neighbors, while the second acts between the $CX$ control qubit and one of its neighbors. All injected $ZZ$ interactions are restricted to qubits within the same register as the $CX$, so that no coupling occurs between registers. 

The second error model, referred to as \textit{inter-register crosstalk}, also injects two $ZZ$ interactions after each $CX$ gate. In this case the interaction couples the $CX$ control (or target) qubit with a qubit belonging to the other register, thereby introducing correlated errors across the two registers.

In both models a fixed two-qubit depolarizing channel is applied after each $CX$ gate and the injected $ZZ$ interactions. This depolarizing channel serves two purposes: it provides a more realistic representation of noise observed on current hardware and establishes a consistent baseline for comparing the effects of intra- and inter-register crosstalk. The $ZZ$ interaction is parameterized by a rotation angle $\theta$, allowing the interaction strength to be tuned. Measurement circuits are kept error-free in order to isolate the effect of crosstalk during random circuit execution.

First, polarization decay remains exponential and depth-dependent when cross-register $ZZ$ errors are injected. For both weak and strong interaction strengths (Fig.~\ref{fig:inter_register_crosstalk_decay}), the characteristic decay structure is preserved, allowing $p_{\mathrm{eff}}$ to be extracted from Eq.~\ref{eq:epcl_fit_func}. This demonstrates that the overlap estimator remains well behaved even in the presence of cross-register correlations

Second, the extracted \ac{epcl} estimates differ between the two error models for the same interaction strength. As shown in Fig.~\ref{fig:inter_and_intra_register_crosstalk_on_peff}, estimates of $p_{\mathrm{eff}}$ obtained using the inter-register model are consistently lower than those obtained using the intra-register model across all interaction strengths. Because errors are injected only during random circuit execution and not in the measurement circuit, this deviation arises from the structure of the injected $ZZ$ interactions. In the intra-register model the injected errors act independently within each register, whereas the inter-register model directly couples qubits belonging to different registers. Such couplings can generate correlations between the two registers and therefore produce contributions beyond the approximate model that are consistent with a non-negligible register covariance term.

Additional evidence supporting this interpretation arises from the behavior of coherent errors under random circuit averaging. As discussed previously, \ac{epcl} is largely insensitive to coherent two-qubit errors unless they are randomized through circuit compilation or twirling, since such errors do not produce the characteristic depth-dependent decay. Consistent with this expectation, intra-register $ZZ$ errors produce little effect unless random compilation is applied. In contrast, inter-register $ZZ$ errors exhibit clear depth-dependent decay even without twirling, indicating that additional mechanisms beyond local coherent error accumulation contribute to the observed behavior.

Finally, inter-register crosstalk exhibits both clear exponential decay with circuit depth (Fig.~\ref{fig:inter_register_crosstalk_decay}) and a monotonic decrease in the extracted polarization with increasing interaction strength (Fig.~\ref{fig:inter_and_intra_register_crosstalk_on_peff}). These results demonstrate that \ac{epcl} is sensitive to this class of cross-register error. However, because the two registers are no longer independent, the interpretation of $p_{\mathrm{eff}}$ derived in the theoretical model no longer strictly applies. In this regime, $p_{\mathrm{eff}}$ should instead be regarded as an effective parameter describing the decay of the overlap estimator, which now reflects contributions from both local errors and the register covariance term $\Delta_{\mathrm{reg}}$.

\subsection{Measurement circuit errors} \label{sec:measurement-error}

Up to this point, our \ac{epcl} simulations have assumed an ideal overlap measurement circuit. In practice, however, the gates used to implement the overlap measurement are subject to the same physical error mechanisms as the benchmarked random circuits. These additional errors can degrade estimator accuracy through both bias and increased variance. Beyond gate-level noise, imperfect qubit readout introduces classical errors that further distort the measured output distribution.

In this section, we study the impact of both measurement-circuit noise and readout error on \ac{epcl}. Single- and two-qubit depolarizing errors are injected into both the random circuit and the overlap measurement circuit following the model introduced in Sec.~\ref{sec:local-depol}. Error strengths span $0 \leq \varepsilon_2 \leq 8.1 \times 10^{-3}$, a regime inclusive of both weak-noise and noise slightly beyond, and $\varepsilon_1 = 0.1 \varepsilon_2$. Readout error is modeled as an independent symmetric bit-flip channel applied to each measured qubit after ideal measurement, with identical probability $p_{\mathrm{meas}}$.

\begin{figure*}[hbt!]
     \centering
     \begin{subfigure}[b]{0.45\textwidth}
         \centering
         \includegraphics[width=\textwidth]{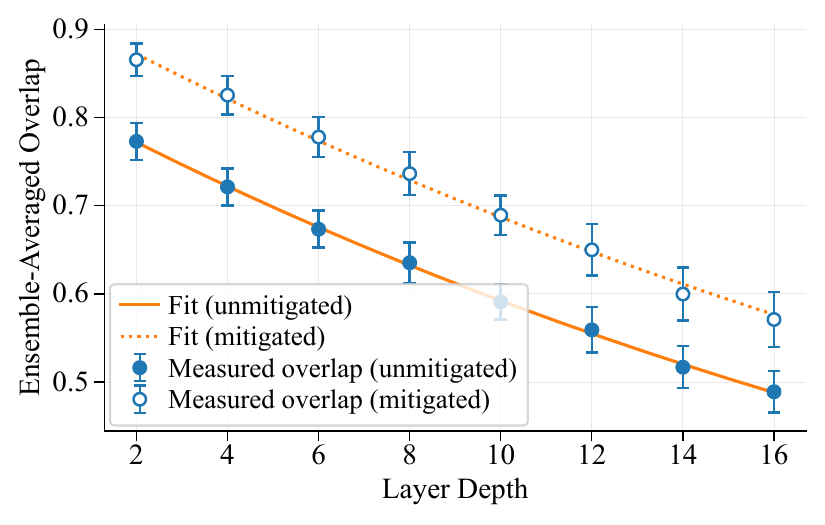}
         \caption{N=8}
         \label{fig:berlin-8q-L20}
     \end{subfigure}
     \begin{subfigure}[b]{0.45\textwidth}
         \centering
         \includegraphics[width=\textwidth]{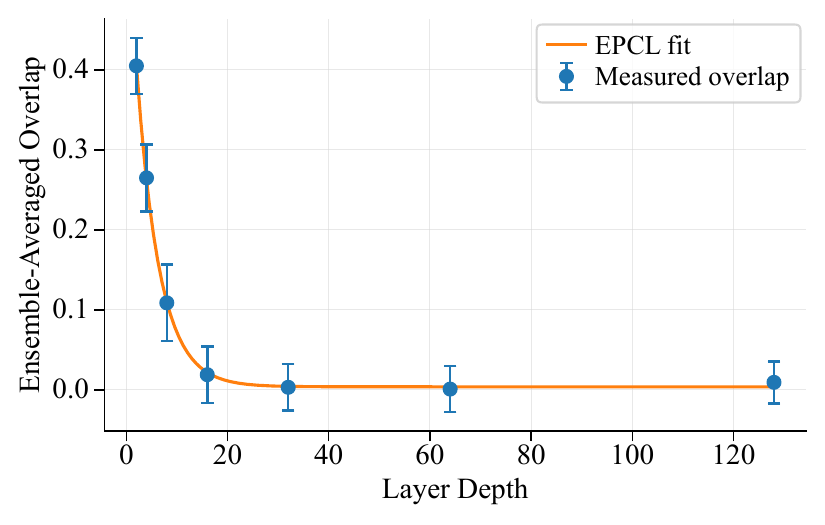}
         \caption{N=16}
         \label{fig:berlin-16q-L20}
     \end{subfigure}
    \caption[EPCL hardware results on IBM Berlin.]
    {Ensemble-averaged \ac{epcl} overlap measured on IBM Berlin as a function of circuit-layer depth.
    (a) Results for $N=8$ benchmark qubits, comprising two $4$-qubit registers, without measurement-error mitigation (filled markers and solid fit) and with measurement-error mitigation (open markers and dotted fit). Both exhibit clear exponential decay. The unmitigated fit gives $p_{\mathrm{eff}}=0.98190\pm1.9\times10^{-4}$ and $A=0.9088\pm1.3\times10^{-3}$, while the mitigated fit gives $p_{\mathrm{eff}}=0.98390\pm4.1\times10^{-4}$ and $A=0.9626\pm2.9\times10^{-3}$.
    (b) Results for $N=16$ benchmark qubits, comprising two $8$-qubit registers, without measurement-error mitigation. The overlap decays more rapidly and approaches the asymptotic value $B=2^{-8}$; the fit gives $p_{\mathrm{eff}}=0.8949\pm 3.0\times 10^{-3}$ and $A=0.7944\pm 7.7 \times 10^{-3}$.
    Lines show fits to the \ac{epcl} overlap-decay model, and error bars indicate the sample standard deviation across the $20$ random circuit instances at each depth.}
    \label{fig:berlin-hardware}
\end{figure*}

We compare the swap-test and Bell-basis overlap measurement circuits for benchmark sizes of $N=4$ and $N=10$ qubits, comprising two $2$-qubit and two $5$-qubit registers, respectively. For each pair of depolarizing strength and readout error probability, we simulate $L=20$ independent \ac{epcl} experiments and estimate the resulting fitted effective polarization parameter. Estimation accuracy is quantified using the \ac{rmse} relative to a high-shot ideal reference computed using an error-free measurement circuit.

To directly compare the two overlap measurement implementations, we analyze the logarithm of the RMSE ratio,
\begin{align}
    \log\!\left(
    \frac{\mathrm{RMSE}_{\mathrm{swap}}}
         {\mathrm{RMSE}_{\mathrm{Bell}}}
    \right),
\end{align}
where positive values indicate lower estimation error for the Bell measurement circuit. The logarithmic scale provides a symmetric visualization of relative performance, preventing extreme ratios from dominating the comparison.

Figure~\ref{fig:measurement_analysis} shows that the Bell measurement circuit achieves lower RMSE than the swap-test implementation across nearly the entire simulated error landscape for both benchmark sizes. This behavior is primarily driven by the substantially greater circuit depth of the swap-test measurement implementation. For the $N=4$ benchmark, the transpiled swap-test measurement circuit requires approximately $16$ $CX$ gates, compared to only $2$ for the Bell measurement circuit. For the $N=10$ benchmark, these counts increase to approximately $40$ and $5$, respectively. As a result, depolarizing gate errors accumulate much more strongly in the swap-test implementation.

Interestingly, this remains true even in regimes where readout error is relatively large. For example, at $p_{\mathrm{meas}} = 4.4 \times 10^{-2}$ and $\varepsilon_2 = 10^{-3}$, one might expect the swap-test measurement to benefit from measuring fewer qubits. Instead, the accumulated gate noise from the substantially deeper swap-test circuit outweighs the readout advantage.

As the benchmark size increases from $N=4$ to $N=10$, the log-RMSE ratio remains positive, indicating that the Bell measurement circuit continues to outperform the swap-test approach. However, the performance gap narrows. This is expected: while the Bell measurement circuit remains substantially shallower than the swap-test implementation, its measured qubit count also increases with benchmark size, making its estimator variance increasingly sensitive to readout error.

These simulations assume all-to-all connectivity, making the reported comparison somewhat conservative. On connectivity-constrained architectures, such as superconducting processors, additional routing operations would increase the depth of both overlap measurement circuits. However, because the swap-test implementation already requires substantially more entangling operations, it would be expected to incur a proportionally larger routing overhead, further increasing its sensitivity to gate-level noise.

\section{Demonstration on IBM Quantum Processor} \label{sec:hardware}

While numerical simulation provides a controlled environment for validating theoretical predictions, the practical utility of a quantum benchmark ultimately depends on its behavior on real hardware. Here we execute \ac{epcl} on IBM's quantum processors to evaluate its behavior under realistic hardware noise and determine whether its characteristic exponential decay and associated $p_{\mathrm{eff}}$ remain observable and interpretable on contemporary quantum processors. 

IBM's superconducting processors provide a particularly useful testbed for evaluating \ac{epcl} under realistic device noise. For these experiments, we select square-lattice qubit configurations that support benchmark circuit structure with minimal routing overhead. This allows the behavior of \ac{epcl} to be studied in the presence of heterogeneous hardware errors while limiting additional error introduced by transpiler-inserted routing operations.

\subsection{System \& Methodology}

Experiments were run on IBM Berlin, a 120-qubit quantum processor in the Nighthawk r1 family of QPUs. This processor was selected due to its large qubit count and comparitavely low, stable error rates at the time of experimentation. 

Qubits are connected in a $12 \times 10$ rectangular lattice. Sections of the processor were selected and benchmarked based on error structure. Benchmark regions were selected using calibration data and supplemental crosstalk characterization, which is quite prominent~\cite{takou_2026}. The specific selection criteria for each experiment are discussed in the corresponding subsections. Calibration and characterization data used at the time of experimentation can be found at \cite{extradata}.

Random circuit construction followed the same method as described in Sec.~\ref{sec:simulation} using $T=4$ and layer depths of $m\in[2, 4, 6, 8, 10, 12, 14, 16]$. At each layer depth the overlap estimate is averaged over $20$ independently sampled random circuits. Each random circuit was independently measured using $2000$ shots. Bell-pair measurement circuits were used throughout due to their lower circuit depth and reduced sensitivity to measurement error, as discussed in Sec.~\ref{sec:measurement-error}.

Randomized compiling was incorporated into the fixed two-qubit layers as described in Sec.~\ref{sec:coherent-errors}. This modification converts coherent over-rotation errors into stochastic fluctuations in the observed decay, allowing \ac{epcl} to remain sensitive to coherent error mechanisms that would otherwise evade detection.

\subsection{Verification of the $N=8$ EPCL benchmark}

\ac{epcl} was first evaluated on an $N=8$ benchmark comprising two $4$-qubit registers, matching the benchmark size used in the numerical simulations. This choice provides a natural comparison point between simulated and hardware behavior. Further, as discussed in Sec.~\ref{sec:measurement-error}, smaller registers are expected to exhibit lower estimator variance due to the reduced number of measured qubits, improving the stability of the resulting fits.

For this initial study, an $8$-qubit region forming a compact $2\times4$ rectangular lattice was selected. At the time of execution, they exhibited relatively low gate-error rates that were broadly within the regime considered in the numerical simulations. Unlike the homogeneous single- and two-qubit depolarizing error rates used in simulation, however, the Berlin processor exhibited substantial spatial variation in both error mechanisms. The compact lattice geometry allowed the benchmark circuits to be implemented with minimal routing overhead, reducing additional error from transpiler-inserted operations.

Figure~\ref{fig:berlin-16q-L20} shows the ensemble-averaged overlap measured on the $N=8$ benchmark. Considering first the unmitigated results shown by the filled markers and solid fit, the expected exponential decay is readily observed, and the measured overlap is well described by the \ac{epcl} overlap-decay model across the investigated depth range. The fit yields $p_{\mathrm{eff}}=0.98190\pm1.9\times10^{-4}$ and $A=0.9088\pm1.3\times10^{-3}$, demonstrating that a stable decay parameter can be extracted despite the presence of realistic hardware noise. The measurement-error-mitigated analysis shown in the same figure is discussed separately in the following subsection.

The experimentally extracted \ac{epcl} of $0.98190\pm1.9\times10^{-4}$ is comparable in magnitude to the values observed in numerical simulation. Direct quantitative comparison is not expected because the hardware exhibits heterogeneous and non-depolarizing noise processes, and the ratio of two-qubit to single-qubit error rates ($\approx 17:1$) differs substantially from the $10:1$ ratio assumed in simulation. Still, the observed value is consistent with the operating regime suggested by the numerical studies and does not indicate any qualitative disagreement between the simulated and experimental behavior of the protocol.

Having established the expected hardware decay behavior, we next examine the sensitivity of the extracted fit parameters to measurement-error mitigation.

\subsection{Sensitivity to measurement-error mitigation}

Measurement-error mitigation is not an intrinsic component of \ac{epcl}, but readout error can alter the measured overlap and may therefore influence the parameters extracted from the decay curve. To assess this sensitivity, the hardware data from the $N=8$ benchmark were also analyzed after applying measurement-error mitigation~\cite{Nation_2021}. These results are shown by the open markers and dotted fit in Fig.~\ref{fig:berlin-8q-L20}.

Mitigation shifts the measured overlap upward across the investigated depth range. The resulting fit yields $p_{\mathrm{eff}}=0.98390\pm4.1\times10^{-4}$ and $A=0.9626\pm2.9\times10^{-3}$, compared with $p_{\mathrm{eff}}=0.98190\pm1.9\times10^{-4}$ and $A=0.9088\pm1.3\times10^{-3}$ without mitigation. The increase in $A$ is consistent with the correction of readout-induced attenuation in the measured overlap. The accompanying change in $p_{\mathrm{eff}}$ indicates that the effect is not limited to a depth-independent change in signal contrast and can also influence the extracted decay rate.

Both analyses nevertheless exhibit a clear exponential decay and produce stable fits. Measurement-error mitigation should therefore be viewed as an optional analysis step rather than a requirement of the \ac{epcl} protocol. Because no independent hardware reference for the underlying layer polarization is available, this comparison does not establish that either fit is the more accurate estimate. Instead, it quantifies the sensitivity of the reported \ac{epcl} parameters to the treatment of measurement error.

These initial hardware demonstrations verified \ac{epcl} at $N=8$ and examined its sensitivity to measurement-error mitigation. We next consider larger qubit subsets, where heterogeneous error accumulation, routing constraints, and estimator variance become increasingly important.

\subsection{Larger-register demonstration}

To evaluate \ac{epcl} with larger individual registers, the demonstration was repeated with an $N=16$ benchmark comprising two $8$-qubit registers. The selected qubits formed a compact $2\times8$ rectangular lattice. The experiment used $20$ independently sampled random circuits at each layer depth $m\in\{2,4,8,16,32,64,128\}$, with $1024$ shots per circuit. No measurement-error mitigation was applied.

Figure~\ref{fig:berlin-hardware}(b) shows the resulting ensemble-averaged overlap. Compared with the smaller-register experiment, the overlap decays more rapidly and approaches the $8$-qubit-register asymptotic value $B=2^{-8}$ by approximately $m=32$. Fitting the complete measured depth range to the \ac{epcl} overlap-decay model gives $p_{\mathrm{eff}}=0.8949\pm 3.0 \times 10^{-3}$ and $A=0.7944\pm 7.7 \times 10^{-3}$, with $B=2^{-8}$ fixed by the register dimension. Despite the faster decay and correspondingly shorter range over which appreciable overlap remains, the measured signal is well described by the same single-exponential decay model used for the smaller-register experiment. This result extends the hardware demonstration of \ac{epcl} to larger registers without requiring a change to the benchmark or fitting procedure.

\subsection{Comparison with EPLG}
\label{sec:eplg-comparison}

To place the \ac{epcl} hardware result alongside an established layer-scale benchmark, we compare the method against \ac{eplg}. \Ac{eplg} constructs a layer fidelity from local simultaneous direct-\ac{rb} experiments, whereas \ac{epcl} directly estimates an effective polarization from the overlap decay produced by repeated register-level circuit layers. Because the two protocols use different circuit constructions and report different primary quantities, the comparison is interpreted as a cross-benchmark consistency check rather than a test of equivalence.

The fitted \ac{epcl} polarization and the reported \ac{eplg} value cannot be compared directly. We therefore place both results on an effective layer-fidelity scale. For \ac{epcl}, the fitted effective polarization is converted to process fidelity for a single $n$-qubit register, with $D=2^n$, according to
\begin{align}
    F_{\mathrm{layer}}^{\mathrm{EPCL}}
    =
    \frac{1+(D^2-1)p_{\mathrm{eff}}}{D^2}.
    \label{eq:epcl-layer-fidelity-eplg-comparison}
\end{align}
For \ac{eplg}, the component process fidelities belonging entirely to registers $A$ and $B$ are combined separately to obtain register-resolved layer fidelities $LF_A$ and $LF_B$. Their geometric mean,
\begin{align}
    F_{\mathrm{layer}}^{\mathrm{EPLG,eff}}
    =
    \sqrt{LF_A LF_B},
\end{align}
is used as the \ac{eplg}-derived effective layer fidelity. This approximately matches the two-register structure of the \ac{epcl} observable while avoiding the per-two-qubit-gate normalization of the native \ac{eplg} metric. The complete procedure is given in Appendix~\ref{app:eplg-comparison}.

We first performed this comparison in simulation using an $N=6$ benchmark comprising two $3$-qubit registers arranged on a $2\times3$ lattice and a homogeneous gate-local depolarizing model. Single-qubit operations were assigned depolarizing strength $\varepsilon_1=4\times10^{-4}$, and two-qubit operations were assigned strength $\varepsilon_2=4\times10^{-3}$. The \ac{epcl} fit gives an effective layer fidelity of $0.9759$, while the geometric mean of the register-resolved \ac{eplg} layer fidelities gives an effective layer fidelity of $0.9872$. The two estimates therefore differ by approximately $0.0113$, or $1.13$ percentage points.

The same comparison was then applied to the matched IBM Berlin hardware experiment. The \ac{epcl} fit gives an effective layer fidelity of $0.9733$, while the \ac{eplg}-derived effective layer fidelity is $0.9592$. The two hardware estimates differ by approximately $0.0141$, or $1.41$ percentage points, with \ac{epcl} reporting the higher fidelity.

\begin{table*}[t]
     \caption[Comparison of EPCL- and EPLG-derived effective layer fidelities.]
    {Comparison of \ac{epcl}- and \ac{eplg}-derived effective layer fidelities in simulation and on IBM Berlin. The \ac{epcl}-derived value is obtained by converting the fitted effective polarization to process fidelity, while the \ac{eplg}-derived value is the geometric mean of the register-resolved layer fidelities, $\sqrt{LF_A LF_B}$. The difference is defined as the \ac{epcl}-derived value minus the \ac{eplg}-derived value.}
    \label{tab:epcl-eplg-comparison}
    \centering
    \begin{tabular}{lccc}
        \hline
        Setting
        & \ac{epcl}-derived LF
        & \ac{eplg}-derived effective LF
        & Difference \\
        \hline
        Depolarizing simulation
        & $0.9759$
        & $0.9872$
        & $-0.0113$ \\
        IBM Berlin
        & $0.9733$
        & $0.9592$
        & $+0.0141$ \\
        \hline
    \end{tabular}
\end{table*}

As established by the preceding simulation studies, \ac{epcl} provides a useful estimate of performance under the controlled depolarizing model considered here. In the present comparison, the \ac{epcl}- and \ac{eplg}-derived layer fidelities differ by approximately $1$--$1.5$ percentage points in both simulation and hardware, although the direction of the difference reverses. Because a discrepancy of comparable magnitude is observed in the controlled simulation, the hardware result alone does not indicate that either protocol is systematically more sensitive to hardware error. Instead, the remaining differences likely reflect the limitations of comparing effective scalar fidelities obtained from distinct randomized circuit constructions.

\section{Discussion}

Under the ensemble-averaged depolarizing model, and when register covariance is negligible, the fitted decay parameter can be interpreted as an effective layer polarization. The local depolarizing simulations in Sec.~\ref{sec:local-depol} validate this interpretation in the weak-noise regime, where the extracted \ac{epcl} value agrees with the first-order gate-level baseline. As noise strength increases, deviations from the first-order model emerge as expected, but the measured overlap remains well described by a single-exponential decay. This distinction is important: disagreement with the first-order approximation outside its regime of validity does not imply failure of the benchmark. Rather, it shows that the fitted layer polarization captures higher-order stochastic error accumulation not retained in the first-order baseline. In this sense, \ac{epcl} continues to estimate an effective layer polarization even when the noise is no longer accurately described by a simple first-order depolarizing model.

The simulations also clarify the error mechanisms to which \ac{epcl} is naturally sensitive. Stochastic depolarizing errors produce the expected depth-dependent decay, while fixed coherent errors in the entangling layer may not. In this case, randomized compiling or twirling is necessary to convert coherent structure into stochastic-like decay that can be detected by the protocol. The resulting value of $p_{\mathrm{eff}}$ should then be interpreted as the effective polarization of the randomized implementation, not necessarily of the bare coherent circuit.

A more substantial change in interpretation occurs when the two registers are no longer independent. In the ideal use case, the joint state of the two-registers factorizes, and the measured overlap reduces to $\mathrm{Tr}(\rho_A \rho_B)$, thus \ac{epcl} can be considered as a measurement device for benchmarking an $n$-qubit register. Inter-register crosstalk breaks this factorization. The measured signal then contains both the marginal overlap of the two reduced register states and a covariance contribution generated by correlations during joint execution. In this sense, the benchmarked object is no longer only the marginal $n$-qubit register, but the attempted independent execution of two registers within the surrounding $2n$-qubit hardware region.

This broader sensitivity is operationally meaningful because unintended correlations between nominally independent registers are themselves an important circuit-level failure mode. In the $ZZ$ crosstalk simulations studied here, increased inter-register coupling produces stronger overlap decay, allowing \ac{epcl} to distinguish register pairs with different levels of this error mechanism. This suggests that \ac{epcl} can be useful not only for estimating register-level polarization, but also for evaluating the isolation of candidate register pairs. The direction and magnitude of the covariance contribution, however, depend on the underlying correlated error mechanism. Other forms of shared or correlated noise may affect the overlap signal differently, and in some cases may not produce a monotonic decrease in the fitted polarization. Thus, the covariance sensitivity should be interpreted as a probe of unintended joint-register behavior, with the specific effect on $p_{\mathrm{eff}}$ determined by the physical error process.

The choice of overlap measurement circuit is also central to practical implementation. Across the simulated regimes considered here, Bell-basis measurements yield lower estimation error than the controlled-SWAP implementation, primarily because they require substantially fewer entangling gates. Although Bell measurements expose the estimator to readout error on more qubits, the reduced circuit depth provides a clear advantage in the studied parameter regime. This motivates the use of Bell-basis measurements in the hardware experiments. However, the gate-error cost of the controlled-SWAP implementation and the readout-error sensitivity of Bell-basis measurements scale differently with register size and hardware error rates. Mapping this tradeoff and identifying regimes in which one measurement implementation becomes preferable to the other is an interesting direction for future work.

The hardware results provide an initial validation that the characteristic \ac{epcl} decay remains observable on contemporary quantum processors. In the smaller-register experiment, using two $4$-qubit registers, the measured overlap exhibits a clear exponential decay and yields a stable estimate of $p_{\mathrm{eff}}$ despite heterogeneous, non-depolarizing hardware noise. The larger-register experiment extends this validation to two $8$-qubit registers, for a total of $16$ physical qubits. Although the overlap decays more rapidly and approaches its dimension-dependent asymptotic value over a shorter depth range, the measured signal remains well described by the same single-exponential decay model. Together, these results demonstrate that the effective-decay model remains experimentally useful across both tested register sizes in the presence of real device noise, routing constraints, and finite sampling.

\Ac{epcl} should be viewed as complementary to existing circuit-level benchmarks rather than as a replacement for them. In particular, \ac{eplg} provides local RB-derived information that can be valuable for diagnosing specific two-qubit layers. \Ac{epcl}, by contrast, directly measures a full-register overlap decay and compresses the aggregate effect of circuit execution into a single parameter. This makes \ac{epcl} useful for assessing register-level performance and for probing error mechanisms whose effects may not be fully captured by products of local benchmark estimates.

These results suggest several directions for future work. Additional studies of inter-register error models would help determine which correlated processes produce monotonic decay, which modify the fitted polarization relative to the independent-register model, and which are weakly visible to the overlap observable. Models that explicitly include register covariance could further separate marginal register polarization from joint-register correlation effects, rather than absorbing both into a single fitted parameter. Together, these extensions would strengthen the use of \ac{epcl} as both a register-level performance metric and a probe of subsystem isolation.

\section{Conclusion}

This work introduces \ac{epcl}, an overlap-based circuit-level benchmark for estimating an effective layer polarization from randomized circuit execution. By applying identical random circuits to two registers and directly measuring their output-state overlap, \ac{epcl} avoids both recovery to a known reference state and classical computation of ideal output probabilities. Under the ensemble-averaged depolarizing model and negligible register covariance, the resulting depth-dependent overlap provides a direct estimate of an effective register-level layer polarization.

Numerical simulations establish the conditions under which this interpretation is useful and clarify important departures from the idealized model. Under local stochastic noise, \ac{epcl} agrees with the perturbative gate-level baseline in the weak-noise regime while continuing to exhibit stable single-exponential decay as higher-order error accumulation becomes important. The simulations further show that fixed coherent errors may require randomization to produce the decay measured by \ac{epcl}, while inter register correlations can modify the overlap through a covariance contribution. Hardware experiments using two $4$-qubit and two $8$-qubit registers demonstrate that the characteristic \ac{epcl} decay remains observable and well described by the benchmark model at both tested register sizes.

\Ac{epcl} should therefore be viewed as a complementary circuit-level benchmark rather than a replacement for locally resolved methods. It sacrifices local error attribution in favor of directly probing the aggregate behavior of a complete repeated circuit layer. This makes \ac{epcl} useful for measuring effective register performance and, with appropriate care in the presence of register correlations, for evaluating the isolation of nominally independent subsystems during simultaneous execution.

\section{Acknowledgement}
We are grateful to Daniel A. Lidar for useful discussions. The authors acknowledge support by the Office of the Director of National Intelligence (ODNI), Intelligence Advanced Research Projects Activity (IARPA), under the Entangled Logical Qubits program through Cooperative Agreement Number W911NF-23-2-0216. For A.V. this material is based upon work supported by, or in part by, the U. S. Army Research Laboratory and the U.S. Army Research Office under contract/grant number W911NF2310255. For T.H, S.M, A.D, J.A, and K.R.B., this material is based upon work supported by, or in part by, the Spectator Qubit Army Research Office MURI (W911NF-18-1-0218) and the National Science Foundation (OSI-2326810).The demonstrations on IBM machines were conducted using IBM Quantum Systems provided through the University of Southern California's IBM Quantum Innovation Center. The views expressed are those of the authors and do not reflect the official policy or position of IBM or the IBM Quantum team.

\bibliography{refs}{}
\bibliographystyle{unsrt}

\appendix

\section{EPCL--EPLG comparison procedure}
\label{app:eplg-comparison}

This appendix describes the conversion used to place the \ac{epcl} and \ac{eplg} results on a common effective layer-process-fidelity scale. The conversion matches the two-register aggregation appearing in the \ac{epcl} decay model, but it does not imply that the two protocols implement identical physical circuit layers.

\subsection{Register-resolved EPLG layer fidelities}

The \ac{eplg} experiment produces fitted \ac{rb} decay parameters for residual one-qubit components and paired two-qubit components. For a component $c$ acting on a Hilbert space of dimension $d_c$, the fitted decay parameter $\alpha_c$ is converted to process fidelity according to
\begin{align}
    F_c
    =
    \frac{1+(d_c^2-1)\alpha_c}{d_c^2},
    \label{eq:eplg-component-process-fidelity}
\end{align}
where $d_c=2$ for a one-qubit component and $d_c=4$ for a two-qubit component.

Let $\mathcal{L}_{\mathrm{in}}$ denote the set of \ac{eplg} layers whose components remain within the two \ac{epcl} registers. Layers containing two-qubit gates that connect registers $A$ and $B$ are excluded. For register $R\in\{A,B\}$, the contribution from \ac{eplg} layer $\ell$ is
\begin{align}
    LF_{\ell,R}
    =
    \prod_{c\in\mathcal{C}_{\ell,R}} F_c,
    \label{eq:eplg-register-sublayer-fidelity}
\end{align}
where $\mathcal{C}_{\ell,R}$ is the set of components in layer $\ell$ that act entirely within register $R$. The register-resolved layer fidelity is then
\begin{align}
    LF_R
    =
    \prod_{\ell\in\mathcal{L}_{\mathrm{in}}}
    LF_{\ell,R}.
    \label{eq:eplg-register-layer-fidelity}
\end{align}

Restricting the construction to within-register components preserves the independent-register structure assumed by the \ac{epcl} model. In the hardware data, including the cross-register \ac{eplg} layer changes the combined layer fidelity from approximately $0.920$ to $0.902$. This effect is therefore non-negligible, but the cross-register layer does not represent circuit content contained in the independently evolved \ac{epcl} registers and is not included in the primary comparison.

\subsection{Conversion to an EPCL-equivalent quantity}

For an $n$-qubit register, let $D=2^n$. Each register-resolved \ac{eplg} layer fidelity is converted to an equivalent depolarizing polarization:
\begin{align}
    p_R^{\mathrm{EPLG}}
    =
    \frac{D^2LF_R-1}{D^2-1},
    \qquad R\in\{A,B\}.
    \label{eq:eplg-register-polarization}
\end{align}

If the effective polarizations of the two registers are $p_A$ and $p_B$, their independent contributions to the overlap decay combine as
\begin{align}
    \left(p_Ap_B\right)^m
    =
    \left(\sqrt{p_Ap_B}\right)^{2m}.
\end{align}
The effective polarization corresponding to the \ac{epcl} fit model is therefore the geometric mean
\begin{align}
    p_{\mathrm{eff}}^{\mathrm{EPLG}}
    =
    \sqrt{
        p_A^{\mathrm{EPLG}}
        p_B^{\mathrm{EPLG}}
    }.
    \label{eq:eplg-geometric-mean-polarization}
\end{align}

Finally, this polarization is converted back to an effective process fidelity:
\begin{align}
    F_{\mathrm{layer}}^{\mathrm{EPLG,eq}}
    =
    \frac{
        1+(D^2-1)p_{\mathrm{eff}}^{\mathrm{EPLG}}
    }{D^2}.
    \label{eq:eplg-equivalent-layer-fidelity}
\end{align}

Equations~\ref{eq:eplg-register-polarization}--\ref{eq:eplg-equivalent-layer-fidelity} define the \ac{eplg}-equivalent values reported in Table~\ref{tab:epcl-eplg-comparison}.

\subsection{Why the native EPLG value is not used}

The native \ac{eplg} metric is defined as
\begin{align}
    \mathrm{EPLG}
    =
    1-LF^{1/n_{2q}},
\end{align}
where $n_{2q}$ is the number of two-qubit gates in the \ac{eplg} layer. Applying the same normalization to \ac{epcl} would require assigning an equivalent $n_{2q}$ to one \ac{epcl} layer. Although the number of entangling gates in the \ac{epcl} circuit is known, this count does not make an \ac{epcl} layer equivalent to the layered construction used by \ac{eplg}, which also contains a different collection of randomized single-qubit operations and a different execution schedule.

The comparison therefore uses layer process fidelity rather than a per-two-qubit-gate error. This avoids introducing an arbitrary gate-count normalization while retaining a common scalar measure of effective register-level performance.

\subsection{Interpretation and limitations}

The conversion above is model-mediated. It maps the outputs of the two protocols onto a common effective depolarizing scale, but it does not remove their operational differences. An \ac{epcl} layer contains repeated Haar-random single-qubit operations followed by its chosen entangling structure. An \ac{eplg} layer fidelity is assembled from simultaneous direct-\ac{rb} experiments using random Clifford operations and interleaved two-qubit gates. Agreement is therefore not expected to be exact, even when both protocols are applied to the same register and error model.

In simulation, the difference between the \ac{epcl} and \ac{eplg}-equivalent values is approximately $1.1$ percentage points. On hardware, the difference is approximately $1.4$ percentage points and has the opposite sign. The comparable magnitudes indicate that the hardware residual cannot, by itself, be attributed to relaxation, idle evolution, crosstalk, or another hardware-specific mechanism. Such effects may still influence either protocol, but the present comparison does not isolate them.

\end{document}